\documentclass[twocolumn,nofootinbib,prd,aps,superscriptaddress,tightenlines,preprintnumbers]{revtex4}

\usepackage{amsmath,amsfonts,amsthm,amscd,amssymb}
\usepackage{cancel}
\usepackage{comment}
\usepackage{slashed}
\usepackage{graphicx}
\usepackage{color}
\usepackage[colorlinks=true
,urlcolor=blue
,anchorcolor=blue
,citecolor=blue
,filecolor=blue
,linkcolor=blue
,menucolor=blue
,linktocpage=true
,pdfproducer=medialab
,pdfa=true
]{hyperref}
\usepackage{cleveref}
\usepackage[utf8]{inputenc}

\newcommand{\nn}{\nonumber \\ }
\def\dbar{{\mathchar'26\mkern-12mu d}}

\begin{document}

\title{The One-Loop Correction to Heavy Dark Matter Annihilation}

\author{Grigory Ovanesyan}
\affiliation{Physics Department, University of Massachusetts Amherst, Amherst, MA}
\author{Nicholas L. Rodd}
\author{Tracy R. Slatyer}
\author{Iain W. Stewart}
\affiliation{Center for Theoretical Physics, Massachusetts Institute of Technology, Cambridge, MA}

\begin{abstract}
We calculate the one-loop corrections to TeV scale dark matter annihilation in a model where the dark matter is described by an SU(2)$_{\rm L}$ triplet of Majorana fermions, such as the wino. We use this framework to determine the high and low-scale $\overline{\rm MS}$ matching coefficients at both the dark matter and weak boson mass scales at one loop. Part of this calculation has previously been performed in the literature numerically; we find our analytic result differs from the earlier work and discuss potential origins of this disagreement. Our result is used to extend the dark matter annihilation rate to NLL$^{\prime}$ (NLL+${\cal O}(\alpha_2)$ corrections) which enables a precise determination of indirect detection signatures in present and upcoming experiments.
\end{abstract}

\preprint{MIT-CTP 4852}
\maketitle

\section{Introduction}
\label{sec:int}

It is now well established that if dark matter (DM) is composed of TeV scale Weakly Interacting Massive Particles (WIMPs) then its present day annihilation rate to produce photons is poorly described by the tree-level amplitude. Correcting this shortcoming is important for determining accurate theoretical predictions for existing and future indirect detection experiments focussing on the TeV mass range, such as H.E.S.S \cite{Hinton:2004eu,Abramowski:2013ax}, HAWC \cite{Sinnis:2004je,Harding:2015bua,Pretz:2015zja}, CTA \cite{Consortium:2010bc}, VERITAS \cite{Weekes:2001pd,Holder:2006gi,Geringer-Sameth:2013cxy}, and MAGIC \cite{FlixMolina:2005hv,Ahnen:2016qkx}.

The origin of the breakdown in the lowest order approximation can be traced to two independent effects. The first of these is the so called Sommerfeld enhancement: the large enhancement in the annihilation cross section when the initial states are subject to a long-range potential. In the case of WIMPs this potential is due to the exchange of electroweak gauge bosons and photons. This effect has been widely studied (see for example \cite{Hisano:2003ec,Hisano:2004ds,Cirelli:2007xd,ArkaniHamed:2008qn,Blum:2016nrz}) and can alter the cross section by as much as several orders of magnitude. The Sommerfeld enhancement is particularly important when the relative velocity of the annihilating DM particles is low, as it is thought to be in the present day Milky Way halo.

The second effect is due to large electroweak Sudakov logarithms of the heavy DM mass, $m_{\chi}$, over the electroweak scale, which enhance loop-level diagrams and cause a breakdown in the usual perturbative expansion. The origin of these large corrections can be traced to the fact that the initial state in the annihilation is not an electroweak gauge singlet, and that a particular $\gamma$ or $Z$ final state is selected, implying that the KLN theorem does not apply \cite{Ciafaloni:2000df,Ciafaloni:1998xg,Ciafaloni:1999ub,Chiu:2009mg}. While the importance of this effect for indirect detection has only been appreciated more recently (see for example \cite{Hryczuk:2011vi,Baumgart:2014vma,Bauer:2014ula,Ovanesyan:2014fwa,Baumgart:2014saa,Baumgart:2015bpa}), it must be accounted for, as it can induce $\mathcal{O}(1)$ changes to the cross section. Hryczuk and Iengo \cite{Hryczuk:2011vi} (hereafter HI) calculated the one-loop correction to the annihilation rate of heavy winos to $\gamma \gamma$ and $\gamma Z$, and found large corrections to the tree-level result, even after including a prescription for the Sommerfeld enhancement. These large corrections are symptomatic of the presence of large logarithms $\ln (2m_{\chi}/m_Z)$ and $\ln (2m_{\chi}/m_W)$, which can generally be resummed using effective field theory (EFT) techniques. This observation has been made by a number of authors who introduced EFTs to study a variety of models and final states. The list includes the case of exclusive annihilation into $\gamma$ or $Z$ final states for the standard fermionic wino \cite{Ovanesyan:2014fwa} and also a scalar version of the wino \cite{Bauer:2014ula}, as well as semi-inclusive annihilation into $\gamma + X$ for the wino \cite{Baumgart:2014vma,Baumgart:2014saa,Baumgart:2015bpa} and higgsino \cite{Baumgart:2015bpa}.

In principle the EFT calculations are systematically improvable to higher order and in a manner where the perturbative expansion is now under control. In order to fully demonstrate perturbative control has been regained, however, it is important to extend these works to higher order. To this end, in this paper we extend the calculation of exclusive annihilation of the wino, which has already been calculated to next-to-leading logarithmic (NLL) accuracy \cite{Ovanesyan:2014fwa}. Doing so includes determining the one-loop correction in the full theory, as already considered in HI. Nonetheless the results in that reference were calculated numerically and are not in the form needed to extend the EFT calculation to higher order. As such, here we revisit that calculation and analytically determine the DM-scale (high-scale) one-loop matching coefficients. We further calculate the electroweak-scale (low-scale) matching at one loop, thereby including the effects of finite gauge boson masses. Taken together these two effects extend the calculation to NLL$^{\prime} =~{\rm NLL}+\mathcal{O}(\alpha_2)$ one-loop corrections, where $\alpha_2 = g_2^2 /4\pi$ and $g_2$ is the SU(2)$_{\rm L}$ coupling. We estimate that our result reduces the perturbative uncertainty from Sudakov effects to $\mathcal{O}(1\%)$, improving on the NLL result where the uncertainty was $\mathcal{O}(5\%)$. Our calculation is complementary to the NLL$^{\prime}$ calculation for the scalar wino considered in \cite{Bauer:2014ula}, and where relevant we have cross checked our work against that reference. In Sec.~\ref{sec:NLL} we outline the EFT setup and review the NLL calculation. Then in Sec.~\ref{sec:1WC} we state the main results of this work, the one-loop high and low-scale matching, leaving the details of their calculation to App.~\ref{app:oneloopfull} and App.~\ref{app:lowscalematching} respectively. Detailed cross checks on the results are provided in App.~\ref{app:consistency} and App.~\ref{app:consistencylow}, whilst lengthy formulae are delayed till App.~\ref{app:PiFunctions}. We compare our analytic results to the numerical ones of HI in Sec.~\ref{sec:Comp} and then conclude in Sec.~\ref{sec:conclusion}.

\newpage

\section{The EFT Framework}
\label{sec:NLL}

We begin by outlining the EFT framework for our calculation, and in doing so review the calculation of heavy DM annihilation to NLL, focussing on the treatment of the large logarithms that were partly responsible for the breakdown in the tree-level approximation. We choose the concrete model of pure wino DM -- the same as used in HI and \cite{Ovanesyan:2014fwa} -- to study these effects. Nevertheless we emphasise the point that the central aim is to quantify the effect of large logarithms which can occur in many models of heavy DM, rather than to better understand this particular model. Ultimately it would be satisfying to extend these results to DM with arbitrary charges under a general gauge group to make the analysis less model specific. This is possible for GeV scale DM indirect detection where the tree-level approximation is generally accurate (see for example \cite{Elor:2015bho,Elor:2015tva}).  Understanding the full range of effects first in a simple model is an important step towards this goal.

The model considered takes the DM to be a wino: an SU(2)$_{\rm L}$ triplet of Majorana fermions. As already highlighted, this is a simple example where both the Sommerfeld enhancement and large logarithms are important. Furthermore this model is of interest in its own right. Neutralino DM is generic in supersymmetric theories \cite{Giudice:1998xp,Randall:1998uk}; models of ``split supersymmetry’’ naturally accommodate wino-like DM close to the weak scale, while the scalar superpartners can be much heavier \cite{Wells:2004di,ArkaniHamed:2004fb,Giudice:2004tc}. DM transforming as an SU(2)$_{\rm L}$ triplet has been studied extensively in the literature, both within split-SUSY scenarios \cite{Arvanitaki:2012ps,ArkaniHamed:2012gw,Hall:2012zp} and more generally \cite{Cohen:2013ama,Cirelli:2007xd,Ciafaloni:2012gs}. The model augments the Standard Model (SM) Lagrangian with
\begin{equation}
\mathcal{L}_{\rm DM} = \frac{1}{2} {\rm Tr} \bar{\chi} \left( i \slashed{D} - M_{\chi} \right) \chi\,.
\label{eq:DMlagrangian}
\end{equation}
We take $M_{\chi} = m_{\chi} \mathbb{I}$, such that in the unbroken theory all the DM fermions have the same mass. After electroweak symmetry breaking, the three states $\chi^{1,2,3}$ break into a Majorana fermion $\chi^0$ and a Dirac fermion $\chi^+$. A small mass difference, $\delta m$, between these states is then generated radiatively, ensuring that $\chi^0$ makes up the observed stable DM. Note, however, that both the charged and neutral states will be included in the EFT.

An effective field theory for this model, NRDM-SCET, was introduced in \cite{Ovanesyan:2014fwa} and used to calculate the rates for the annihilation processes $\chi \chi \to ZZ, Z\gamma, \gamma \gamma$. Specifically the EFT generalizes soft-collinear effective theory (SCET) \cite{Bauer:2000ew,Bauer:2000yr,Bauer:2001ct,Bauer:2001yt} to include non-relativistic dark matter (NRDM) in the initial state. Schematically the calculation involves several steps. Firstly the full theory has to be matched onto the relevant NRDM-SCET$_{\rm EW}$ operators at the high scale of $\mu \simeq 2 m_{\chi}$. The qualifier EW indicates that this is a theory where electroweak degrees of freedom -- the $W$ and $Z$ bosons, top quark, and the Higgs -- are dynamical, as introduced in \cite{Chiu:2007yn,Chiu:2007dg,Chiu:2008vv,Chiu:2009mg,Chiu:2009ft}. These operators then need to be run down to the electroweak scale, $\mu \simeq m_Z$. At this low scale, we then match NRDM-SCET$_{\rm EW}$ onto a theory where the electroweak degrees of freedom are no longer dynamical, NRDM-SCET$_{\gamma}$. This matching accounts for the effects of electroweak symmetry breaking, such as the finite gauge boson masses. At this stage we can now calculate the low-scale matrix elements which provide the Sommerfeld enhancement. We now briefly review each of these steps.

The first requirement is to match NRDM-SCET$_{\rm EW}$ and the full theory at the high scale $\mu_{m_{\chi}}$. The relevant operators in the EFT to describe DM annihilation have the following form:
\begin{equation}
O_r = \frac{1}{2} \left( \chi_v^{aT} i \sigma_2 \chi_v^b \right) \left( S_r^{abcd} \mathcal{B}_{n \perp}^{ic} \mathcal{B}_{\bar{n} \perp}^{jd} \right) i \epsilon^{ijk} (n - \bar{n})^k\,,
\label{eq:Ops}
\end{equation}
which is written in terms of the basic building blocks of the effective theory, and in the centre of momentum frame we can define $v=(1,0,0,0)$, $n=(1,\hat{n})$, and $\bar{n}=(1,-\hat{n})$ where $\hat{n}$ is the direction of an outgoing gauge boson. In more detail $\chi_v^a$ is a non-relativistic two-component fermionic field of gauge index $a$ corresponding to the DM and $\mathcal{B}_{\bar{n},n}$ contain the outgoing (anti-)collinear gauge bosons $A_{\bar{n},n}^{\mu}$, which can be seen as
\begin{equation}
\mathcal{B}_{n \perp}^{\mu} = A^{\mu}_{n \perp} - \frac{k_{\perp}^{\mu}}{\bar{n} \cdot k} \bar{n} \cdot A_n^{\mu} + \ldots \,,
\label{eq:SCETB}
\end{equation}
where the higher order terms in this expression involve two or more collinear gauge fields. For $\mathcal{B}_{\bar{n} \perp}^{\mu}$ we simply interchange $n \leftrightarrow \bar{n}$. The full form of $\mathcal{B}_{n \perp}^{\mu}$ can be found in \cite{Bauer:2002nz}, and is collinear gauge invariant on its own. Finally the gauge index connection is encoded in $S_r^{abcd}$:
\begin{equation}\begin{aligned}
S_1^{abcd} &= \delta^{ab} ( \mathcal{S}_n^{ce} \mathcal{S}_{\bar{n}}^{de} )\,,\\
S_2^{abcd} &= ( \mathcal{S}_v^{ae} \mathcal{S}_{n}^{ce} ) ( \mathcal{S}_v^{bf} \mathcal{S}_{\bar{n}}^{df} )\,.
\label{eq:GaugeIndex}
\end{aligned}\end{equation}
These expressions are written in terms of adjoint Wilson lines of soft gauge bosons along some direction $n$, $\bar{n}$, or $v$; in position space the incoming Wilson line is
\begin{equation}
\mathcal{S}_v(x) = P \exp \left[ ig \int_{-\infty}^0 ds v \cdot A_v(x+ns) \right]\,,
\label{eq:SCETS}
\end{equation}
where the matrix $A^{bc}_v = -i f^{abc} A^a_v$ and for outgoing Wilson lines the integral runs from $0$ to $\infty$.

The fact there are only two possible forms of $S_r^{abcd}$ means there are only two relevant NRDM-SCET operators. An important requirement of the operators is that the incoming DM fields must be in an $s$-wave configuration. Then being a two-particle state of identical fermions, the initial state must be a spin singlet. If the annihilation was $p$-wave or higher, it would be suppressed by powers of the low DM velocity relative to these operators. The Wilson coefficients associated with these operators are determined by the matching. Calculating to NLL only requires the tree-level result where $C_1(\mu_{m_{\chi}}) = - C_2(\mu_{m_{\chi}}) = - \pi \alpha_2(\mu_{m_{\chi}})/m_{\chi}$ as an initial condition. Here again $\alpha_2$ is the SU(2)$_{\rm L}$ fine structure constant. We extend this result to one loop in Sec.~\ref{sec:1WC}.

After matching, the next step is to evolve these operators down to the low scale, effectively resumming the large logarithms $\ln(2m_\chi/m_Z)$ and $\ln(2m_\chi/m_W)$ that caused a breakdown in the perturbative expansion of the coupling. This is done using the anomalous dimension matrix $\hat{\gamma}$ of the two operators (a matrix as the operators will in general mix during the running). In general the matrix can be broken into a diagonal piece $\gamma_{W_T}$, and a non-diagonal soft contribution $\hat{\gamma}_S$, as
\begin{equation}
\hat{\gamma} = 2 \gamma_{W_T} \mathbb{I} + \hat{\gamma}_S\,.
\label{eq:anomdim}
\end{equation}
To NLL these results are given by \cite{Ovanesyan:2014fwa}:
\begin{equation}\begin{aligned}
\gamma_{W_T} &= \frac{\alpha_2}{4\pi} \Gamma_0^g \ln \frac{2m_{\chi}}{\mu} - \frac{\alpha_2}{4\pi} b_0 + \left( \frac{\alpha_2}{4\pi} \right)^2 \Gamma_1^g \ln \frac{2m_{\chi}}{\mu}\,, \\
\hat{\gamma}_S &= \frac{\alpha_2}{\pi} (1-i\pi) \begin{pmatrix} 2 & 1 \\ 0 & -1 \end{pmatrix} - \frac{2\alpha_2}{\pi} \begin{pmatrix} 1 & 0 \\ 0 & 1 \end{pmatrix}\,.
\label{eq:anomdimparts}
\end{aligned}\end{equation}
Here the diagonal anomalous dimension has been written in terms of the SU(2)$_{\rm L}$ one-loop $\beta$-function, $b_0 = 19/6$, as well as the cusp anomalous dimensions, $\Gamma_0^g = 8$ and $\Gamma_1^g = 8 \left(\frac{70}{9} - \frac{2}{3} \pi^2 \right)$, and we use the full SM particle content for this evolution.\footnote{This means we take $m_t\sim m_H\sim m_{W,Z}$ and integrate out all these particles at the same time at the electroweak scale.} Renormalization group evolution with the anomalous dimension also requires the two-loop $\beta$-function, and for this we take $b_1 = -35/6$. Our normalization convention is such that $\mu d\alpha_2/d\mu = -b_0 \alpha_2^2/(2\pi) - b_1 \alpha_2^3/(8\pi^2)$.  Below the DM matching scale, the spin of the DM is no longer important. As such the anomalous dimension determined in \cite{Ovanesyan:2014fwa} for the fermionic wino should resum the same logarithms as those that appear in the scalar case considered in \cite{Bauer:2014ula}, and we have confirmed they agree.

We can then explicitly use the full anomalous dimension to evolve the operators as follows:
\begin{eqnarray}
\begin{bmatrix} C_{\pm}^X (\left\{ m_i \right\}) \vspace{0.1cm}\\ C_{0}^X (\left\{ m_i \right\}) \end{bmatrix} &= &e^{\hat{D}^X(\mu_Z,\left\{ m_i \right\}))} P \exp \left( \int_{\mu_{m_{\chi}}}^{\mu_Z} \frac{d\mu}{\mu} \hat{\gamma}(\mu, m_{\chi}) \right) \nonumber \\
&\times& \begin{bmatrix} C_1(\mu_{m_{\chi}},m_{\chi}) \\ C_2(\mu_{m_{\chi}},m_{\chi}) \end{bmatrix}\,, \label{eq:Running}
\end{eqnarray}
Let us carefully explain the origin and dependence of each of these terms. Starting from the right, $C_1$ and $C_2$ are the high-scale Wilson coefficients of the operators stated in Eq.~\eqref{eq:Ops}, resulting from a matching of the full theory onto NRDM-SCET$_{\rm EW}$. These only depend on the high scales, specifically $\mu_{m_{\chi}}$ and $m_{\chi}$. Next the anomalous dimension $\hat{\gamma}$ is also a high scale object, and so only depends on $m_{\chi}$ and now $\mu$ as it runs between the relevant scales. $\hat{D}^X$ is a factor accounting for the low-scale matching from NRDM-SCET$_{\rm EW}$ onto NRDM-SCET$_{\gamma}$ -- a theory where the electroweak modes have been integrated out, see \cite{Chiu:2007yn,Chiu:2007dg,Chiu:2008vv,Chiu:2009mg,Chiu:2009ft}. It is a matrix as soft gauge boson exchanges can mix the operators. Furthermore $\hat{D}^X$ is labelled by $X$ to denote its dependence on the specific final state considered, $\gamma \gamma$, $\gamma Z$ or $ZZ$. This object depends on the low-scale physics and so depends on $\mu_Z$ and all the masses in the problem, which we denote as $\{ m_i \}$. It contains both a resummation of low-scale logarithms (which can be carried out directly as in~\cite{Chiu:2007yn,Chiu:2007dg} or more elegantly with the rapidity renormalization group~\cite{Chiu:2012ir}, see also~\cite{Becher:2010tm}) as well as the low scale matching coefficient which does not necessarily exponentiate.  Finally on the left we have our final coefficients $C_{\pm}^X$ and $C_{0}^X$, which as explained below can be associated with the charged and neutral annihilation processes. In an all orders calculation of all terms in Eq.~\eqref{eq:Running}, the scale dependence would completely cancel on the right hand side, implying that $C_{\pm}^X$ and $C_{0}^X$ depend only on the mass scales in the problem and not $\mu_{m_{\chi}}$ or $\mu_Z$. Nevertheless at any finite perturbative order, the scale dependence does not cancel completely and so a residual dependence is induced in these coefficients. We will exploit this to estimate the uncertainty in our results associated with missing higher order terms.

As we are performing a resummed calculation, the order to which we calculate is defined in terms of the large electroweak logarithms we can resum. In general the structure of the logarithms can be written schematically as:
\begin{equation}
\ln \frac{C}{C^{\rm tree}} \sim \sum_{k=1}^{\infty} \left[ \vphantom{\alpha_2^k \ln^{k+1}} \right. \underbrace{\alpha_2^k \ln^{k+1}}_{\rm LL} + \underbrace{\alpha_2^k \ln^k}_{\rm NLL} + \underbrace{\alpha_2^k \ln^{k-1}}_{\rm NNLL} + \ldots \left. \vphantom{\alpha_2^k \ln^{k+1}} \right]\,,
\label{eq:NLLprimeDef}
\end{equation}
where since Sudakov logarithms exponentiate, we have defined the counting in terms of the log of the result. Furthermore all corrections are defined with respect to the tree level result $C^{\rm tree} \sim \mathcal{O}(\alpha_2)$, which is a convention we will follow throughout. With this definition of the counting, to perform the running in Eq.~\eqref{eq:Running} to NLL order, there are three effects that must be accounted for: 1. high-scale matching at tree level; 2. two-loop cusp and one-loop non-cusp anomalous dimensions; and 3. the low-scale matching at tree level, together with the rapidity renormalization group at NLL. To extend this to NNLL all three of these need to be calculated to one order higher. In between these two is the NLL$^{\prime}$ result we present here, which involves determining both the high and low-scale matching at one loop. In terms of Eq.~\eqref{eq:NLLprimeDef}, this amounts to determining the leading $k=1$ piece of the NNLL result. To the extent that $\mathcal{O}(\alpha_2)$ corrections are larger than those at $\mathcal{O}(\alpha^2_2 \ln (\mu_{m_{\chi}}^2/\mu_Z^2))$, the NLL$^{\prime}$ result is an improvement over NLL and more important than NNLL.

Before presenting the result of that calculation, however, it is worth emphasising another advantage gained from the effective theory. In addition to allowing us to resum the Sudakov logarithms, the effective theory also allows this problem to be cleanly separated from the issue of low-velocity Sommerfeld enhancement in the amplitude -- in NRDM-SCET there is a Sommerfeld-Sudakov factorization. At leading power the relevant SCET Lagrangian contains no interaction with the DM field. On the other hand NRDM does contain soft modes, which are responsible for running the couplings, however these modes do not couple the Sommerfeld potential to the hard interaction at leading power. Consequently matrix elements for the DM factorize from the matrix elements of the states annihilated into. This allows for an all orders factorized formula for the DM annihilation amplitude in this theory:
\begin{equation}\begin{aligned}
\mathcal{M}_{\chi^0\chi^0 \to X} &= 4 \sqrt{2} m_{\chi} P_X \left[ s_{00} \left( \Sigma_1^X - \Sigma_2^X \right) + \sqrt{2}  s_{0\pm} \Sigma_1^X \right]\,, \\
\mathcal{M}_{\chi^+\chi^- \to X} &= 4 m_{\chi} P_X \left[ s_{\pm0} \left( \Sigma_1^X - \Sigma_2^X \right) + \sqrt{2}  s_{\pm\pm} \Sigma_1^X \right]\,.
\label{eq:Factorized}
\end{aligned}\end{equation}
Here $X$ can be $\gamma \gamma$, $\gamma Z$ or $ZZ$ and $P_{\gamma \gamma} = - e^2 \epsilon_{n \perp}^i \epsilon_{\bar{n} \perp}^j \epsilon^{ijk} \hat{n}^k/(2m_{\chi})$, whilst $P_{\gamma Z} = \cot \bar{\theta}_W P_{\gamma \gamma}$ and $P_{ZZ} = \cot^2 \bar{\theta}_W P_{\gamma \gamma}$, with $\bar{\theta}_W$ the $\overline{\rm MS}$ Weinberg angle. The key physics in this equation is that the contribution from Sommerfeld enhancement is captured in the terms $s_{ij}$, whilst the contribution from electroweak logarithms is in $\Sigma_i^X$; the two are manifestly factorized and can be calculated independently.

The focus of the present work is to extend the calculation of the Sudakov effects. In terms of the factorized result stated in Eq.~\eqref{eq:Factorized} this amounts to an improved calculation of $\Sigma_i^X$. Explicitly, from there we can see that:
\begin{equation}\begin{aligned}
\left| \Sigma_1^X \right|^2 &= \frac{\sigma_{\chi^+ \chi^- \to X}^{\cancel{\rm SE}}}{\sigma_{\chi^+ \chi^- \to X}^{\rm tree}}\,, \\
\left| \Sigma_1^X - \Sigma_2^X \right|^2 &= \frac{\sigma_{\chi^0 \chi^0 \to X}^{\cancel{\rm SE}}}{\sigma_{\chi^+ \chi^- \to X}^{\rm tree}}\,,
\label{eq:SigmaDef}
\end{aligned}\end{equation}
where $\cancel{\rm SE}$ denotes a calculation where Sommerfeld Enhancement is intentionally left out. To be even more explicit, we can write these Sudakov effects in terms of the Wilson coefficients in Eq.~\eqref{eq:Running}. Specifically we have:
\begin{equation}\begin{aligned}
\Sigma_1^X &= \frac{C_{\pm}^X}{C_1^{\rm tree}}\,, \\
\Sigma_1^X - \Sigma_2^X &= \frac{C_{0}^X}{C_1^{\rm tree}}\,,
\label{eq:SigmaDefExplicit}
\end{aligned}\end{equation}
where as stated above $C_1^{\rm tree} = - \pi \alpha_2/m_{\chi}$.

\section{The One-Loop Correction}
\label{sec:1WC}

In this section we discuss the main results of this work, which includes analytic expressions for both the high and low scale matching coefficients in the language introduced in the previous section. We start with reporting the result of the calculation of the high-scale Wilson coefficients $C_r$ to one loop. The details have been eschewed to App.~\ref{app:oneloopfull}. In short this calculation involves enumerating and evaluating the 25 one-loop diagrams that mediate $\chi^a \chi^b \to W^c W^d$ in the unbroken full theory and then matching this result onto the NRDM-SCET$_{\rm EW}$ operators. For example, we evaluate diagrams such as
\begin{center}
\includegraphics[height=0.15\columnwidth]{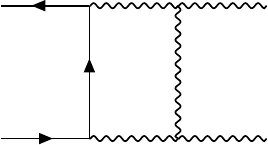} \hspace{0.1in}
\includegraphics[height=0.15\columnwidth]{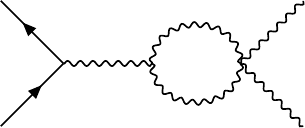}
\end{center}
and provide the analytic expression graph by graph. Here the solid lines are DM particles and wavy lines are electroweak gauge bosons in the full theory above the DM scale. In addition we account for the counter term contribution, the change in the running of the coupling through the matching, and also ensure that the calculation maintains the Sudakov-Sommerfeld factorization. Combining all of these we find 
\begin{equation}\begin{aligned}
C_1(\mu) &=- \frac{\pi\alpha_2(\mu)}{m_{\chi}} + \frac{\alpha_2(\mu)^2}{4m_{\chi}} \left[ 2 \ln^2 \frac{\mu^2}{4m_{\chi}^2} \right. \\
&\left. + 2 \ln \frac{\mu^2}{4m_{\chi}^2} + 2 i \pi \ln \frac{\mu^2}{4m_{\chi}^2} + 12 - \frac{11\pi^2}{6} \right] \,, \\
C_2(\mu) &=\frac{\pi\alpha_2(\mu)}{m_{\chi}} - \frac{\alpha_2(\mu)^2}{2m_{\chi}} \left[ \ln^2 \frac{\mu^2}{4m_{\chi}^2} \right. \\
&\hspace{0.218in}\left. + 3 \ln \frac{\mu^2}{4m_{\chi}^2} - i \pi \ln \frac{\mu^2}{4m_{\chi}^2} + 2 - \frac{5\pi^2}{12} \right]\,.
\label{eq:WilsonCoeff}
\end{aligned}\end{equation}
Here and throughout this section $\alpha_2(\mu)$ is the coupling defined below the scale of the DM mass, $m_{\chi}$. We explain this distinction carefully in App.~\ref{app:oneloopfull}. For each coefficient in Eq.~\eqref{eq:WilsonCoeff} the first term represents the tree-level contribution. A cross check on this result is provided in App.~\ref{app:consistency}, where we check that the $\mu$ dependence of this result properly cancels with that of the NLL resummation from~\cite{Ovanesyan:2014fwa} for the $\mathcal{O}(\alpha_2)$ corrections. The cancellation occurs between our result in Eq.~\eqref{eq:WilsonCoeff} and the running induced by the anomalous dimension stated in Eqs.~\eqref{eq:anomdim} and \eqref{eq:anomdimparts}; this can be seen clearly in Eq.~\eqref{eq:Running} as these are the only two objects that depend on $\mu_{m_{\chi}}$. As the anomalous dimension is independent of the DM spin, the logarithms appearing in our high-scale matching coefficients should also be, and indeed ours match those in the scalar calculation of \cite{Bauer:2014ula}. Of course the finite terms should not be, and are not, the same.

We next state the contribution from the low-scale matching. Unsurprisingly, as this effect accounts for electroweak symmetry breaking effects such as the gauge boson masses, it is in general dependent upon the identity of the final states. Again this is a matching calculation and involves evaluating diagrams that appear in SCET$_{\rm EW}$, but not SCET$_{\gamma}$, and we provide three examples below.
\begin{center}
\includegraphics[height=0.2\columnwidth]{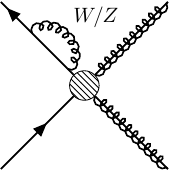} \hspace{0.1in}
\includegraphics[height=0.2\columnwidth]{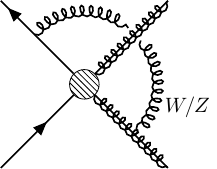} \hspace{-0.05 in}
\includegraphics[height=0.2\columnwidth]{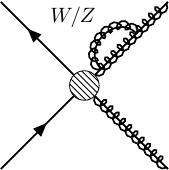}
\end{center}
Here springs with a line through them are collinear gauge bosons with energy $\sim m_\chi$ in the DM center-of-mass frame, and springs without the extra line are soft gauge bosons with energy $\sim m_{W,Z}$.  A central difficulty in the calculation is accounting for the effects of electroweak symmetry breaking, see for example \cite{Denner:2016etu} for a recent discussion. In order to simplify this we make use of the general formalism for electroweak SCET of \cite{Chiu:2007yn,Chiu:2007dg,Chiu:2008vv,Chiu:2009mg,Chiu:2009ft}, which we have extended to include the case of non-relativistic external states.\footnote{This calculation can also be performed using the rapidity renormalization group \cite{Chiu:2012ir}, but in order to make best use of earlier SCET calculations in SCET$_{\rm EW}$ we will not use that formalism here.} We postpone the details to App.~\ref{app:lowscalematching}. The approach breaks the full low-scale matching into a ``soft" and ``collinear" component, which are the labels associated with the non-diagonal and diagonal contributions respectively, rather than the effective theory modes that give rise to them. This distinction is discussed further in App.~\ref{app:lowscalematching}. In our case, $\hat{D}^X(\mu)$ in Eq.~\eqref{eq:Running} can be specified through
\begin{equation}
\exp \left[\hat{D}^X(\mu) \right] = \left[ \hat{D}_s(\mu) \right] \left[ \vphantom{\hat{D}} D_c^{\chi}(\mu) \mathbb{I} \right] \exp \left[ \sum_{i\in X} D_c^i(\mu) \mathbb{I} \right] \,,
\label{eq:lowbreakdown}
\end{equation}
where again $X$ can be $\gamma \gamma$, $\gamma Z$ or $ZZ$, $\hat{D}_s(\mu)$ is the non-diagonal soft contribution and a matrix as it mixes the operators, whilst $D_c^{\chi}(\mu)$ and $D_c^i(\mu)$ are the initial and final state diagonal contributions respectively. Note both $\hat{D}_s(\mu)$ and the identity matrix $\mathbb{I}$ are $2 \times 2$ matrices. The terms that are not exponentiated in Eq.~\eqref{eq:lowbreakdown} are only determined to $\mathcal{O}(\alpha_2)$, whereas the final state diagonal contribution has its logarithmically enhanced contribution resummed to all orders. Using this definition we find that the components of the soft matrix are (see App.~\ref{app:lowscalematching}):
\begin{eqnarray}
~{[\hat{D}_s]}_{11} &=& 1 + \frac{\alpha_2(\mu)}{2\pi} \left[ \ln \frac{m_W^2}{\mu^2} (1-2i\pi) + c_W^2 \ln \frac{m_Z^2}{\mu^2} \right]\,, \nn
{[\hat{D}_s]}_{12} &=& \frac{\alpha_2(\mu)}{2\pi} \ln \frac{m_W^2}{\mu^2} (1-i\pi)\,, \label{eq:LowSoft} \\
{[\hat{D}_s]}_{21} &=& 1 + \frac{\alpha_2(\mu)}{2\pi} \ln \frac{m_W^2}{\mu^2} (2-2i\pi)\,, \nn
{[\hat{D}_s]}_{22} &=& 1\,. \nonumber
\end{eqnarray}
Here and throughout we use the shorthand $c_W = \cos \bar{\theta}_W$ and $s_W = \sin \bar{\theta}_W$. Further, the diagonal contributions can be written as:
\begin{equation}\begin{aligned}
D_c^{\chi}(\mu) = & 1 - \frac{\alpha_2(\mu)}{2\pi} \left[ \ln \frac{m_W^2}{\mu^2} + c_W^2 \ln \frac{m_Z^2}{\mu^2} \right]\,, \\
D_c^i(\mu) = &\frac{\alpha_2(\mu)}{2\pi} \left[ \ln \frac{m_W^2}{\mu^2} \ln \frac{4 m_{\chi}^2}{\mu^2} - \frac{1}{2} \ln^2 \frac{m_W^2}{\mu^2} \right. \\
&\hspace{0.5in}\left. - \ln \frac{m_W^2}{\mu^2} + c^i_1 \ln \frac{m_Z^2}{\mu^2} + c^i_2 \right]\,,
\label{eq:LowColinear}
\end{aligned}\end{equation}
where $i = Z$ or $\gamma$ and we have:
\begin{equation}\begin{aligned}
c^Z_1 &= \frac{5-24s_W^2-22s_W^4}{24c_W^2}\,, \\
c^{\gamma}_1 &= 1-\frac{47}{36}s_W^2\,,
\label{eq:LowColinearConsts1}
\end{aligned}\end{equation}
and 
\begin{equation}\begin{aligned}
c^Z_2 &= -1.5534 - 3.0892 i\,, \\
c^{\gamma}_2 &= -0.812092\,.
\label{eq:LowColinearConsts2}
\end{aligned}\end{equation}
Analytic expressions for these last results are provided in App.~\ref{app:lowscalematching} and App.~\ref{app:PiFunctions}, and we give numerical values here as the expressions are lengthy. Note that we have distinguished between factors of $m_W$ and $m_Z$ in all logarithms. 

The $\mu$ dependence of the low-scale matching is demonstrated to cancel with that in our high-scale matching result when the running is turned off, the details being shown in App.~\ref{app:consistencylow}. We emphasise that this cross check involves not only the $\mu$ dependence of the objects in Eq.~\eqref{eq:lowbreakdown}, but also the $\mu$ dependence of the high-scale coefficients stated in Eq.~\eqref{eq:WilsonCoeff} and further the SM SU(2)$_{\rm L}$ and U(1)$_Y$ $\beta$-functions. The full $\mu$ cancellation is non-trivial -- it requires the interplay between each of these objects. This ultimately provides us with confidence in the results as stated. As a further check, our low-scale matching result does not depend on the spin of the DM. As such we should be again able to compare our result to the scalar case calculated in \cite{Bauer:2014ula}. In that work they only considered the $\gamma \gamma$ final state, and also neglected the impact of SM fermions. Restricting our calculation to the same assumptions, we confirm that the $\mu$ dependence in our result matches theirs.

\begin{figure*}[t!]
\centering
\begin{tabular}{c}
\includegraphics[scale=0.37]{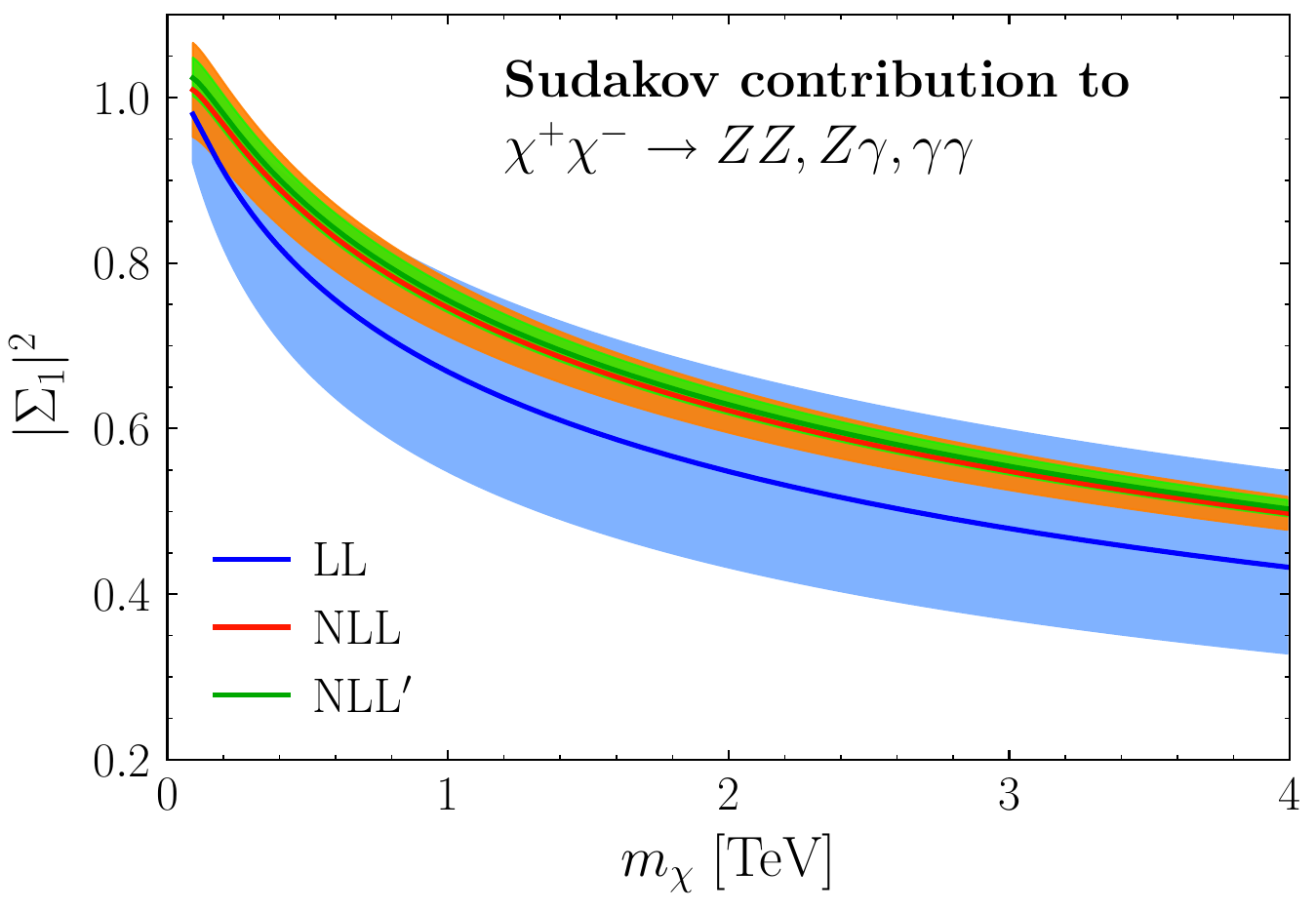} 
\hspace{0.12in}
\includegraphics[scale=0.378]{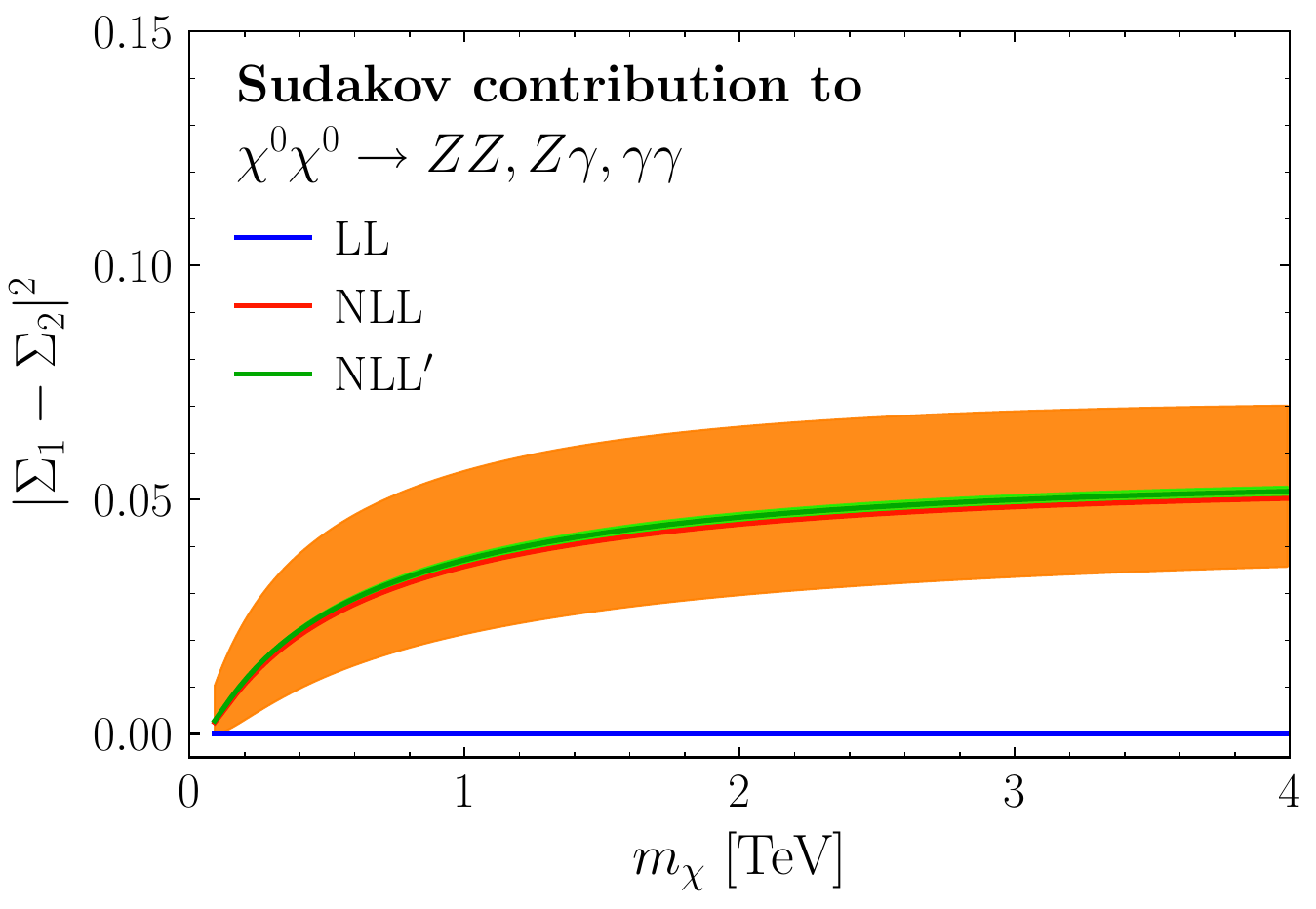}
\end{tabular}
\caption{\footnotesize{Here we show our NLL$^{\prime}$ result for the electroweak corrections to the charged (left) and neutral (right) DM annihilations obtained by adding the one-loop high and low-scale corrections to the NLL result. The result is in good agreement with the known NLL calculation, but with smaller uncertainty since the scale uncertainties have been reduced. The bands here are derived by varying the high scale between $m_{\chi}$ and $4 m_{\chi}$.}}
\label{fig:Nll1Loop}
\end{figure*}

\begin{figure*}[t!]
\centering
\begin{tabular}{c}
\includegraphics[scale=0.37]{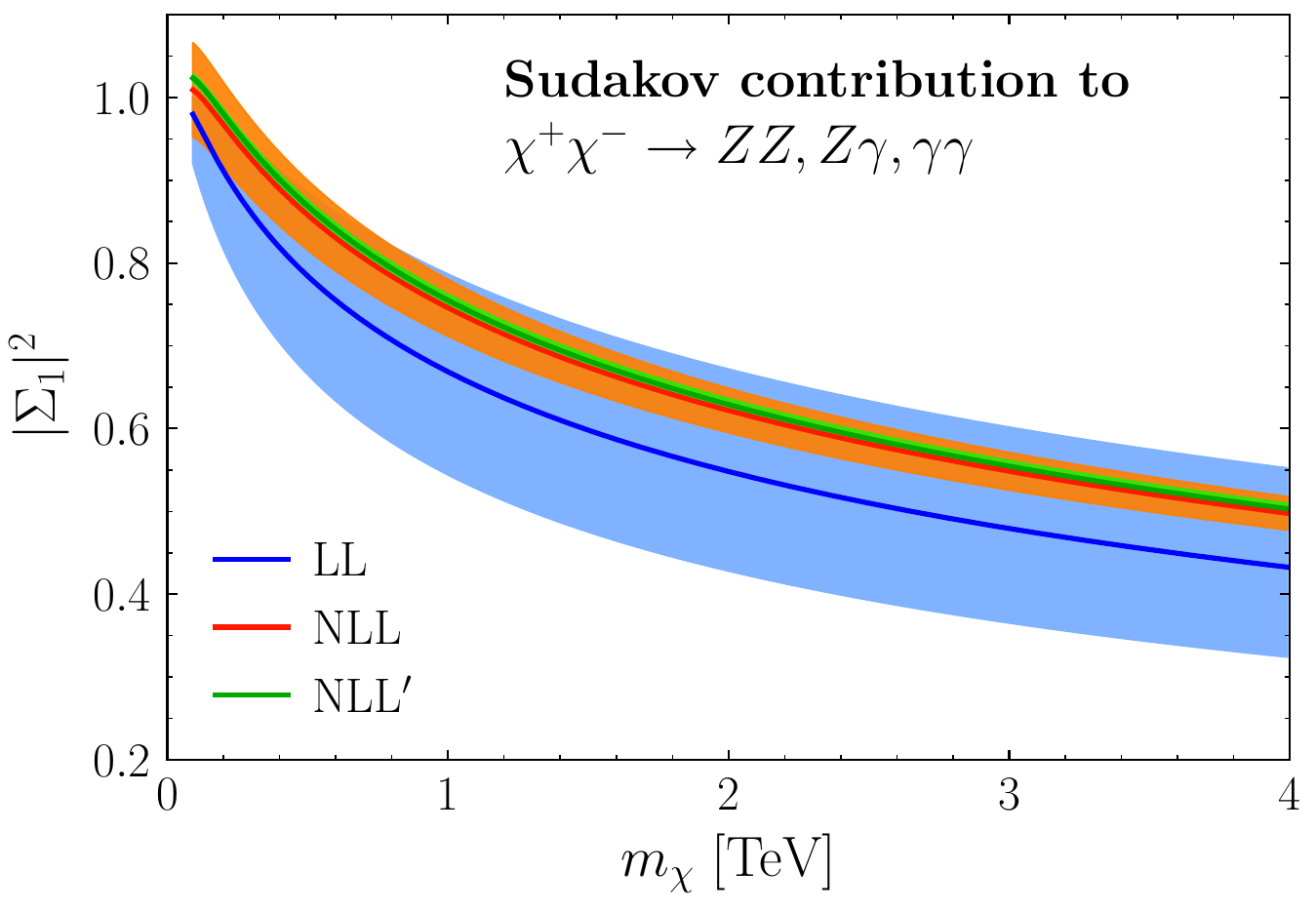} \hspace{0.12in}
\includegraphics[scale=0.378]{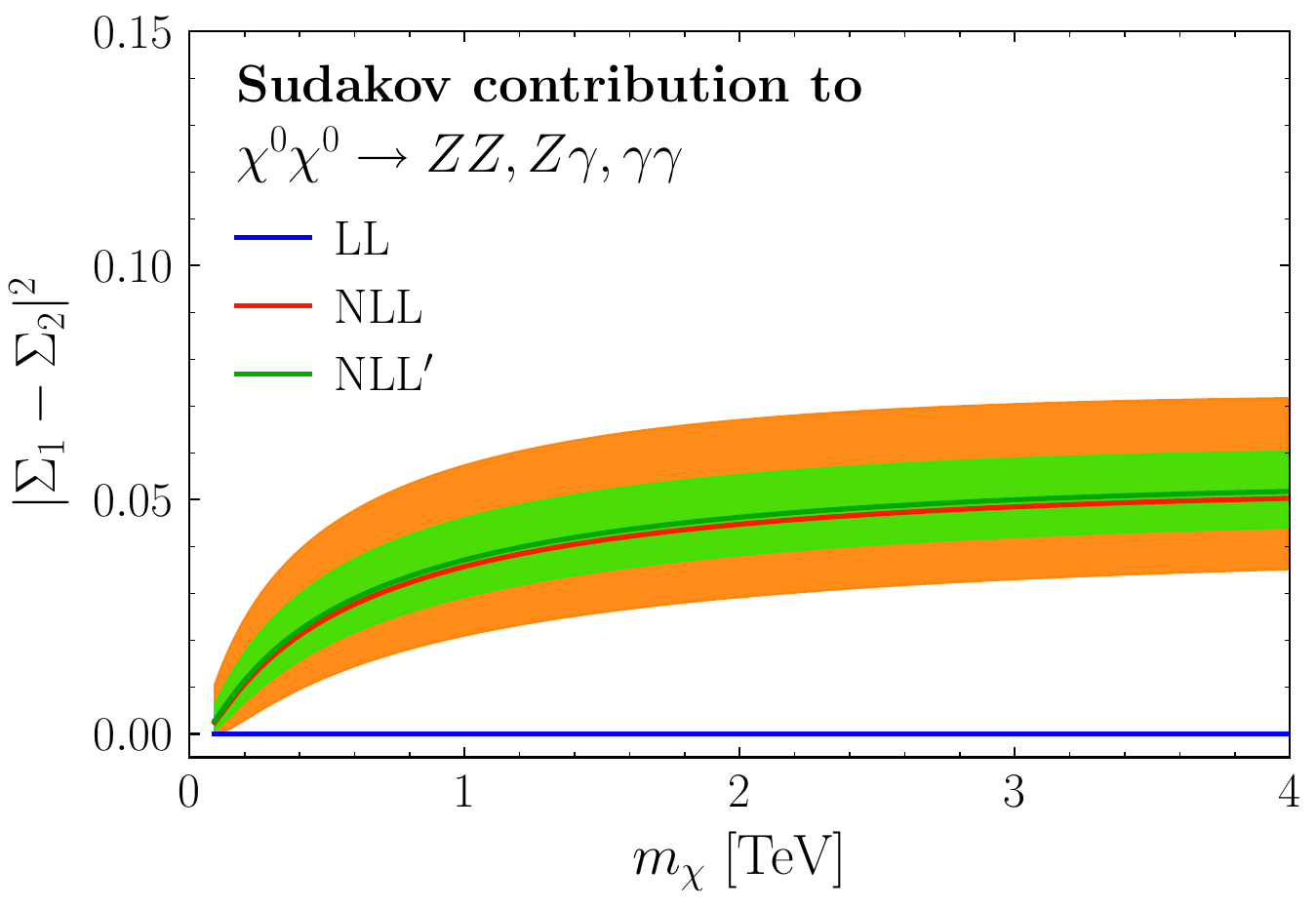}
\end{tabular}
\caption{\footnotesize{As for Fig.~\ref{fig:Nll1Loop}, but showing a variation in the low-scale matching between $m_Z/2$ and $2m_Z$, rather than a variation of the high-scale matching. As can be seen the NLL$^{\prime}$ contribution has reduced the low scale dependence in both charged and neutral DM annihilation cases, and is again consistent with the NLL result.}}
\label{fig:Nll1LoopLowScale}
\end{figure*}

\begin{figure}[t!]
\centering
\begin{tabular}{c}
\includegraphics[scale=0.37]{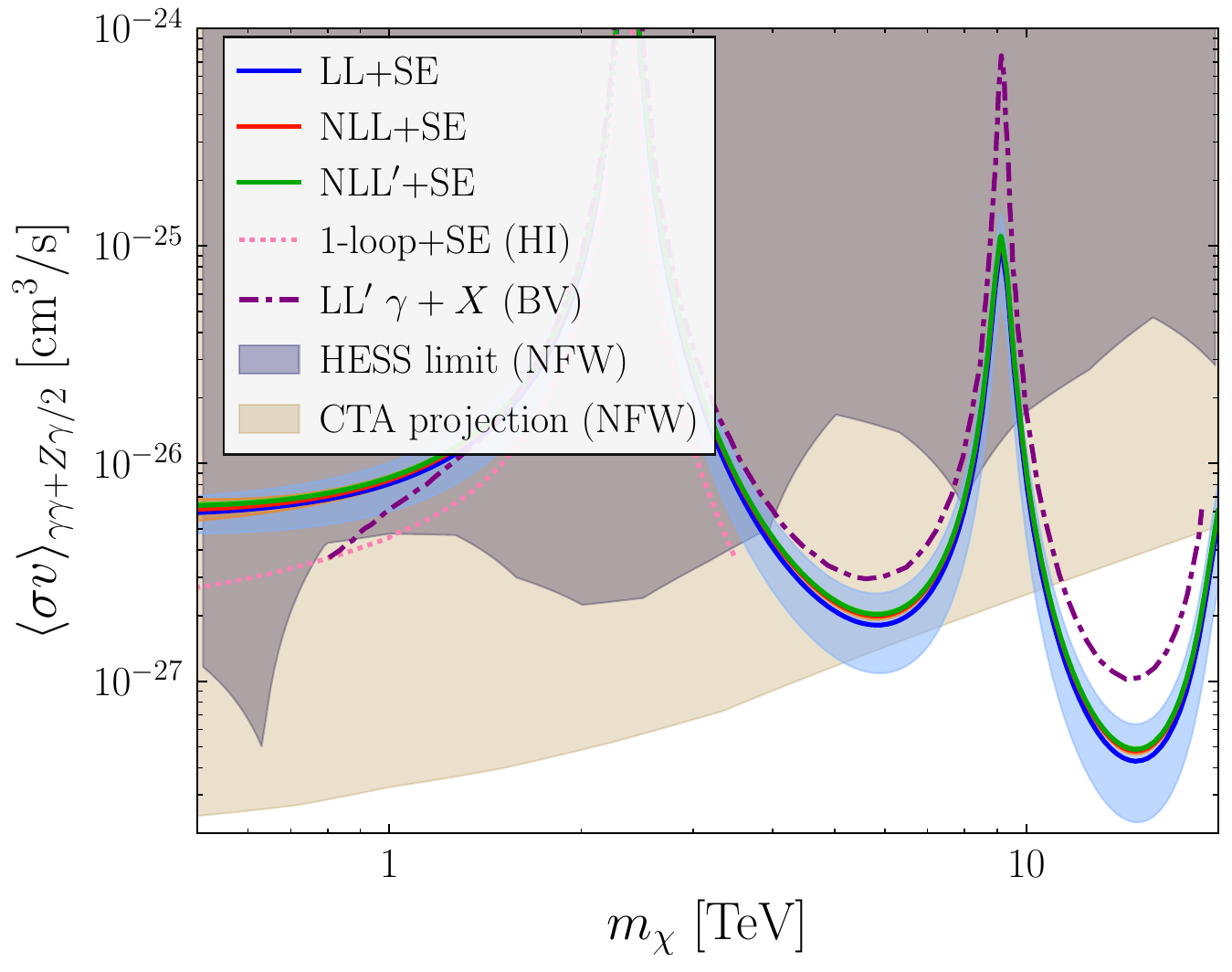}
\end{tabular}
\caption{\footnotesize{The impact of the NLL$^{\prime}$ result on the full cross section, which includes the Sommerfeld Enhancement (SE), is shown to be consistent with the lower orders result, suggesting the electroweak corrections are under control. Also shown is the rate for the semi-inclusive process $\gamma + X$ calculated to LL$^{\prime}$ in \cite{Baumgart:2015bpa}. In addition on this plot we show current bounds from H.E.S.S. and projected ones from CTA, determined assuming 5 hours of observation time. See text for details.}}
\label{fig:CrossSec}
\end{figure}

Taking our results in combination, we can extend the NLL calculation to NLL$^{\prime}$. Of course we cannot show full NNLL results in the absence of the higher order anomalous dimension calculation, nevertheless the results we state here determine the cross section with perturbative uncertainties on the Sudakov effects reduced to the percent level. At $\mathcal{O}(\alpha_2^2)$, our calculation accounts\footnote{Again note that all counting here is relative to the lowest order contribution, which occurs at $C^{\rm tree} \sim \mathcal{O}(\alpha_2)$. As such the absolute order of the terms in this sentence is $\mathcal{O}(\alpha_2^3)$.} for all terms of the form $\alpha_2^2 \ln^4 (\mu_{m_{\chi}}^2/\mu_Z^2)$, $\alpha_2^2 \ln^3 (\mu_{m_{\chi}}^2/\mu_Z^2)$, and $\alpha_2^2 \ln^2 (\mu_{m_{\chi}}^2/\mu_Z^2)$. The first perturbative term we are missing at this order is $\alpha_2^2 \ln (\mu_{m_{\chi}}^2/\mu_Z^2)$. Taking $\mu_Z = m_Z$ and $m_{\chi}$ anywhere from $m_Z$ to $20$ TeV, we estimate the absence of these terms induces an uncertainty that is less than 1\%, demonstrating the claimed accuracy.

To combine the various results stated above into the cross section we take the factorized results in Eq.~\eqref{eq:Factorized}, and note that as the higher order Wilson coefficients have nothing to do with the Sommerfeld enhancement, their contribution is included in the $\Sigma$ terms as given explicitly in Eq.~\eqref{eq:SigmaDefExplicit}. We know that at tree level $s_{00} = s_{\pm \pm} = 1$ and $s_{0 \pm} = s_{\pm 0}=0$, implying that when the Sommerfeld enhancement can be ignored we can associate $|\Sigma_1|^2$ with the Sudakov contribution to $\chi^+ \chi^-$ annihilation and $|\Sigma_1 - \Sigma_2|^2$ with $\chi^0 \chi^0$.

For this reason, in Fig.~\ref{fig:Nll1Loop} and Fig.~\ref{fig:Nll1LoopLowScale} we show the contributions to $|\Sigma_1|^2$ and $|\Sigma_1 - \Sigma_2|^2$ for LL, NLL and NLL$^{\prime}$. In both cases we see the addition of the one-loop corrections is completely consistent with the NLL results, suggesting that this approach has the Sudakov logarithms under control. In these plots we take a central value of $\mu_{m_{\chi}} = 2 m_{\chi}$ and $\mu_Z = m_Z$. In Fig.~\ref{fig:Nll1Loop} the bands are derived from varying the high-scale matching between $m_{\chi}$ and $4 m_{\chi}$. Recall that if we were able to calculate these quantities to all orders, they would be independent of $\mu$, and so varying these scales estimates the impact of missing higher order terms. For the $|\Sigma_1|^2$ NLL result, taking $\mu_{m_{\chi}} = 2 m_{\chi}$ is a minimum in the range varied over, so we symmetrise the uncertainties in order to more conservatively estimate the range of uncertainty. Similarly in Fig.~\ref{fig:Nll1LoopLowScale} we show the equivalent plot, but here the bands are derived by varying the low scale $\mu_Z$ from $m_Z/2$ to $2m_Z$. Improving on the high and low-scale matching, as we have done here, should lead to a reduction in the scale uncertainty. In all four cases shown this is clearly visible and furthermore all results are still consistent with the NLL result within the uncertainty bands.

\begin{figure*}[t!]
\centering
\begin{tabular}{c}
\includegraphics[scale=0.37]{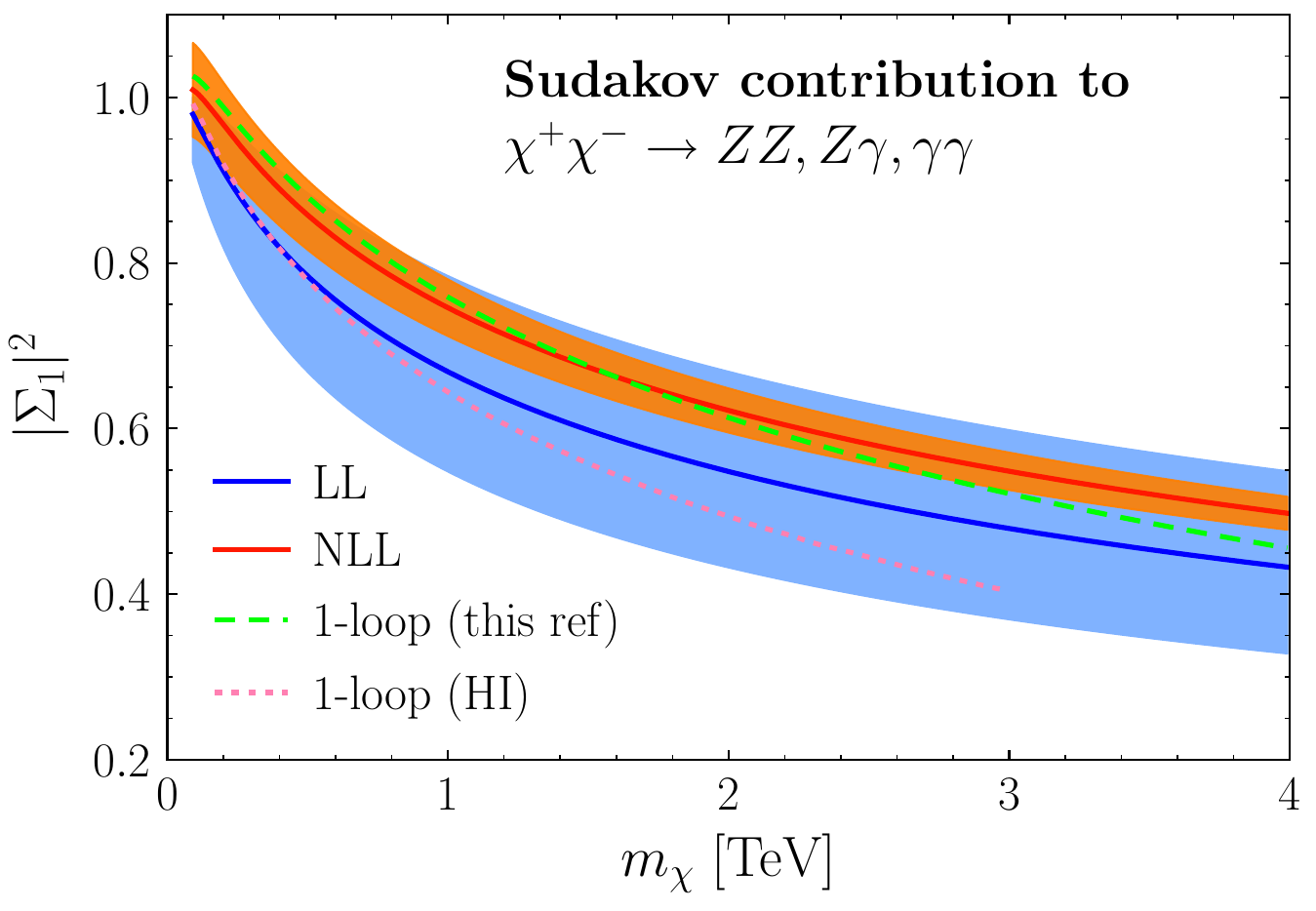} \hspace{0.12in}
\includegraphics[scale=0.378]{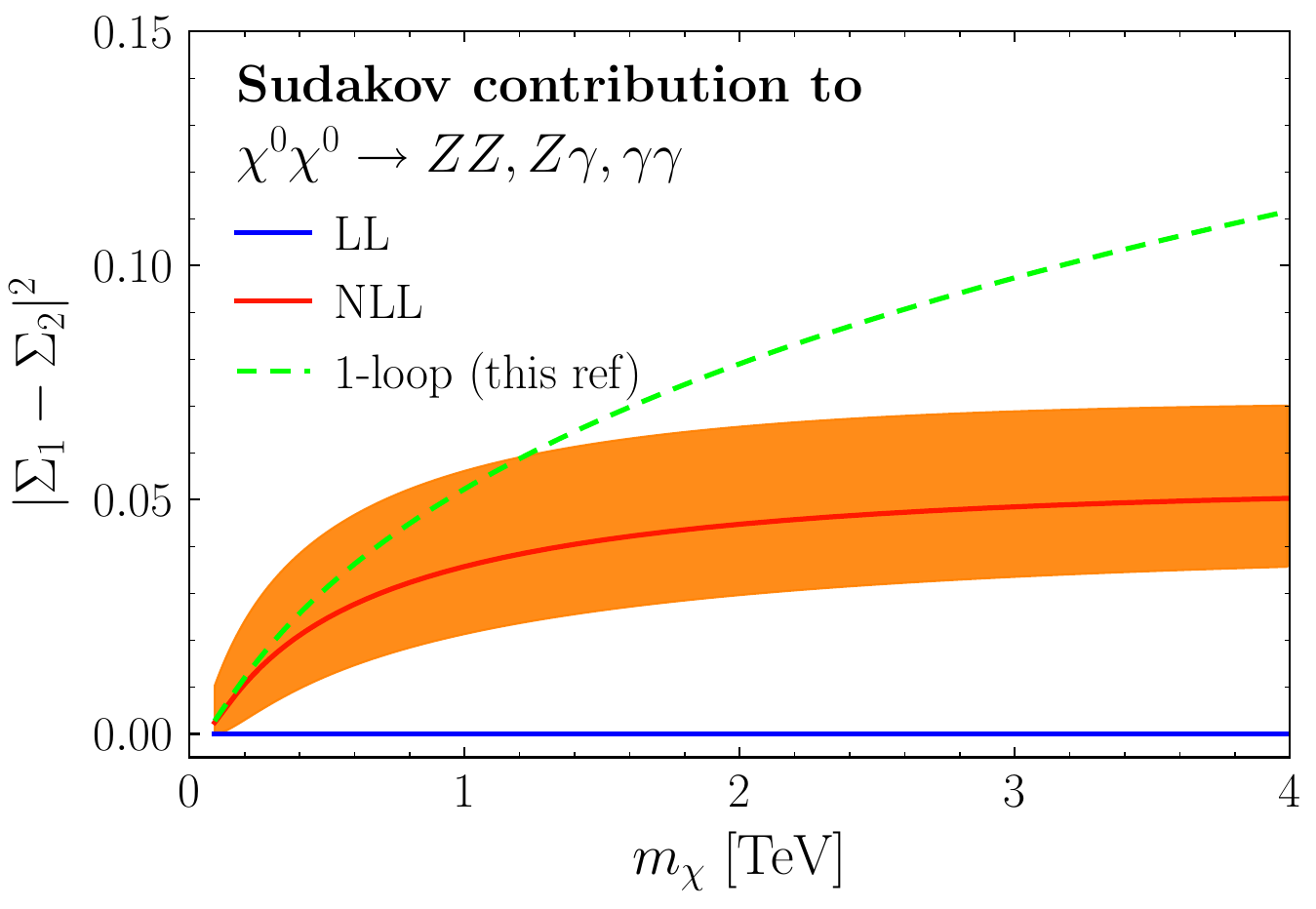}
\end{tabular}
\caption{\footnotesize{Similar to Fig.~\ref{fig:Nll1Loop}, but instead of displaying NLL$^{\prime}$ curves we show our high and low-scale one-loop results including no running from the anomalous dimension. For the case of $\chi^+ \chi^-$ annihilation we further show the equivalent result of HI, taken from Fig.~11 of their work (which only extends up to 3 TeV). There is evidently some discrepancy between the results. Note that at low masses where the Sudakov logarithms are not too large, our result is consistent with the NLL result as would be expected. See text for details.}}
\label{fig:Nllcf1loop}
\end{figure*}

\begin{figure*}[t!]
\centering
\begin{tabular}{c}
\includegraphics[scale=0.37]{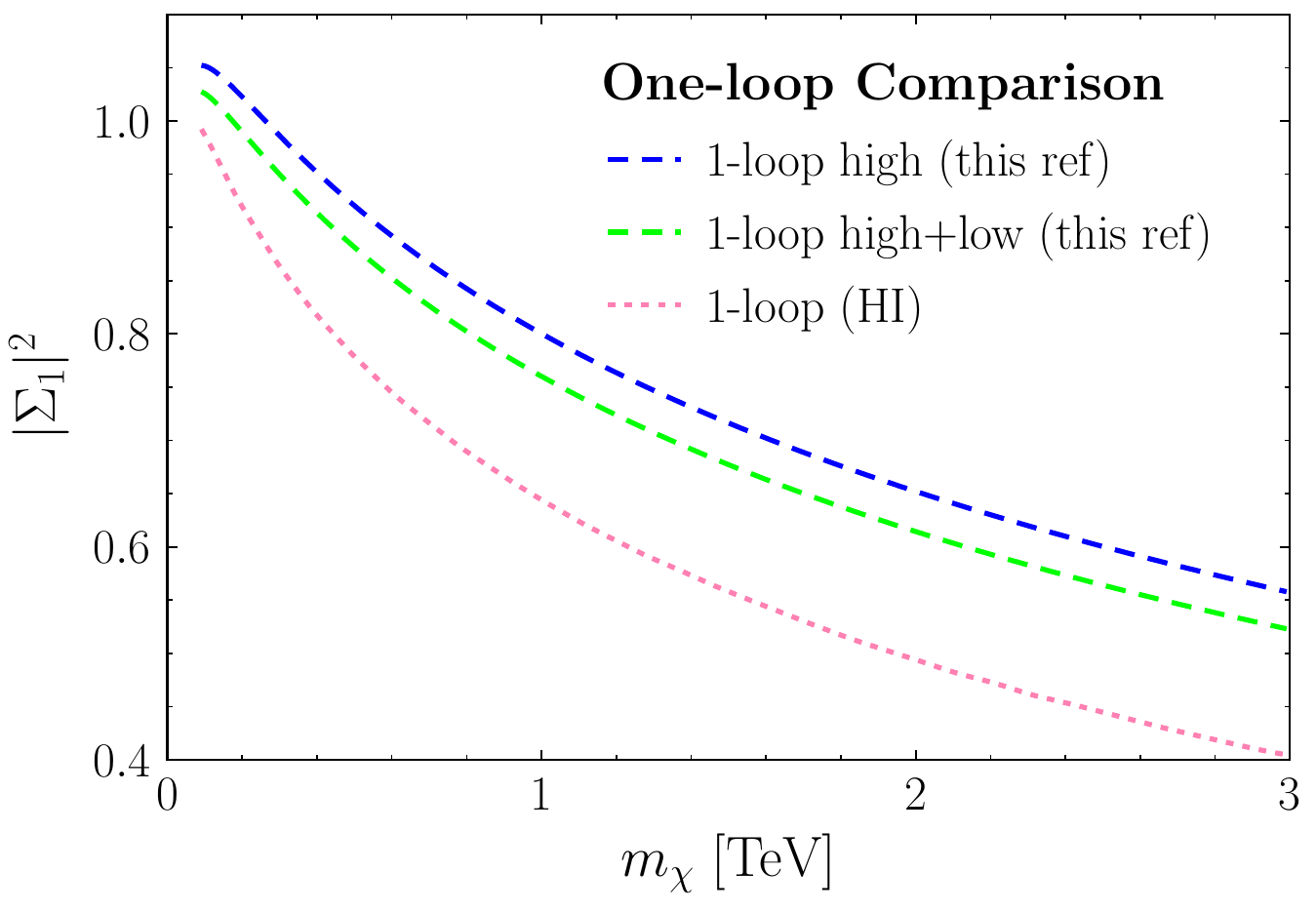} \hspace{0.12in}
\includegraphics[scale=0.37]{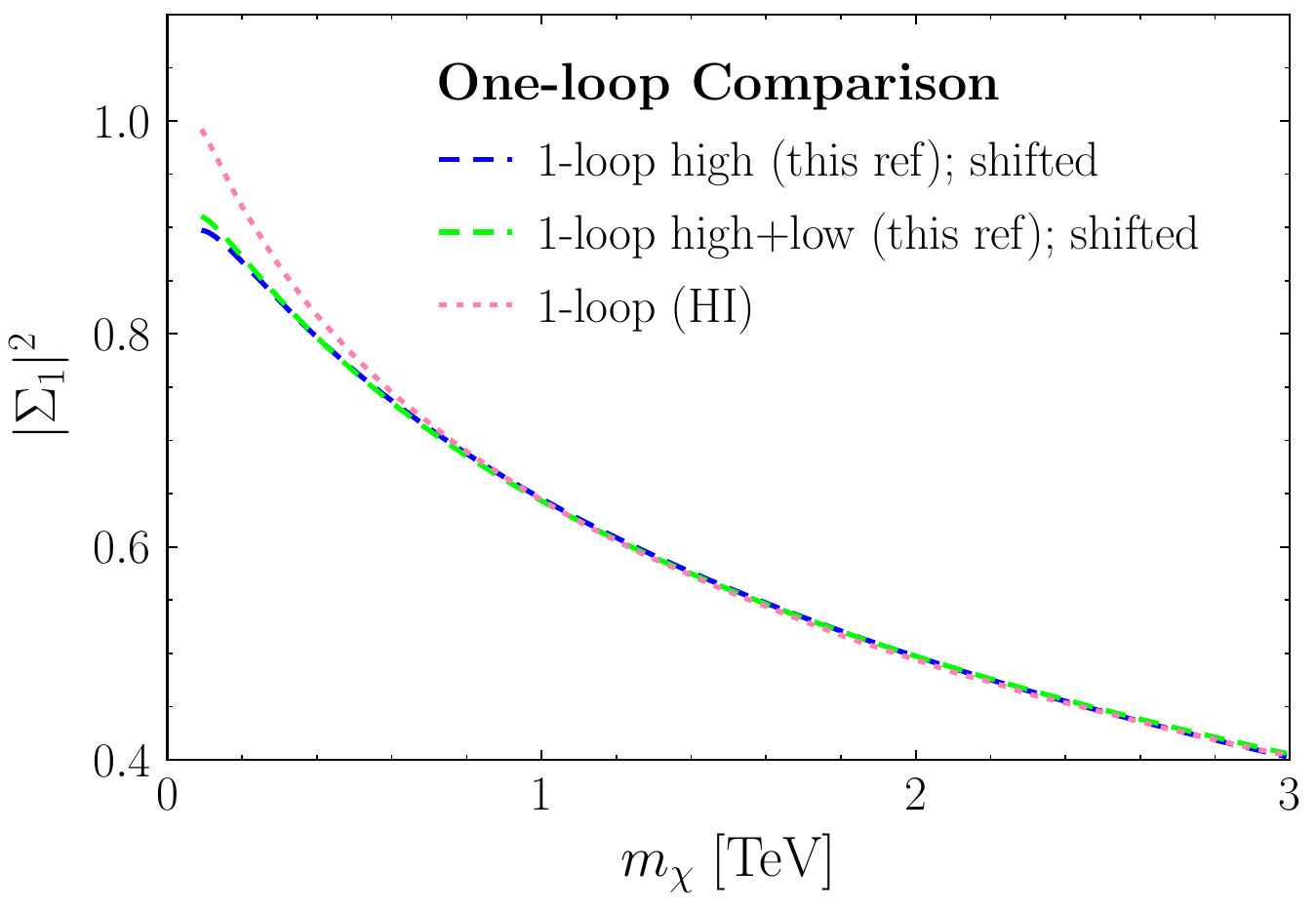}
\end{tabular}
\caption{\footnotesize{We show the result of HI for $|\Sigma_1|^2$ compared to two variations of our result. Firstly in the left panel we show our result with the high only or high and low-scale calculations compared to the result of HI, taken from Fig.~11 of their paper, demonstrating that there is a disagreement. In the right panel we take our results and shift them each by a $m_\chi$ independent constant. The shifted results show that above $\sim$ 1~TeV the $m_\chi$ dependence of our result is in good agreement with HI.}}
\label{fig:UsVsHI}
\end{figure*}

We can also take this result and determine the impact on the full DM annihilation cross section into line photons from $\gamma \gamma$ and $\gamma Z$ in this model, as we show in Fig.~\ref{fig:CrossSec}. We take the uncertainty on our final result to include the high and low-scale variations added in quadrature. For H.E.S.S. limits we use \cite{Abramowski:2013ax}, whilst for the CTA projection we assume 5 hours of observation time and use \cite{Cohen:2013ama,Bergstrom:2012vd}. For both we assume an NFW profile with a local DM density of 0.4 GeV/cm$^3$. We see again that our partial NLL$^{\prime}$ results are consistent with the NLL conclusions.\footnote{A digitized version of our cross section is available with the arXiv submission or upon request.} In this figure we also include the LL$^{\prime}$ result for the semi-inclusive process $\gamma+X$ taken from Fig.~7 of \cite{Baumgart:2015bpa}, denoted by (BV). The semi-inclusive result is above our line photon result, except at low DM masses. Note that this work does not show scale uncertainties, so the precise difference is hard to quantify numerically.

\section{Comparison to Earlier Work}
\label{sec:Comp}

In addition to using our results from the previous section in conjunction with the running due to the anomalous dimension, we can also consider the case where we take our one-loop result in isolation. In this sense we should be able to reproduce the initial problem of large logarithms seen by HI. We show this in Fig.~\ref{fig:Nllcf1loop}, compared to the LL and NLL result. For $\Sigma_1$ our one-loop result is consistent with that from NLL, indicating the importance of the $\alpha_2 \ln^2 (\mu_{m_{\chi}}^2/\mu_Z^2)$ and $\alpha_2 \ln (\mu_{m_{\chi}}^2/\mu_Z^2)$ corrections to $C^{\rm tree}$. For $\Sigma_1 - \Sigma_2$, which starts at NLL, our one-loop result is only consistent with the NLL expression in the small $m_{\chi}$ region.

For the $|\Sigma_1|^2$ case we also show on that plot the equivalent curve for HI as extracted from Fig.~11 of their paper. From here it is clear that the qualitative shape of our results agrees with theirs but that there is disagreement in the normalization. This disagreement is already clear in Fig.~\ref{fig:CrossSec} and is more evident in Fig.~\ref{fig:Nllcf1loop}. In Fig.~\ref{fig:UsVsHI} we analyze this difference in more detail. In the left panel we show the difference between their result and ours, showing our calculation with and without the low-scale matching included. Given the low-scale matching accounts for the electroweak masses, which were included in HI, we would expect including it to improve the agreement. This is seen, but it does not substantially relieve the tension.

To further explore the difference, in the right panel of Fig.~\ref{fig:UsVsHI} we take our results and shift them down by a constant: 0.155 for the high only result and 0.117 for the high and low combination. Such a constant offset could originate from a difference in $m_{\chi}$ independent terms between our result and HI. Unfortunately, however, a difference in such terms could originate from almost any of the graphs contributing to the result. Comparing our analytic expressions to the numerical results of HI we have been unable to pinpoint the exact location of the disagreement, although it is clear that we agree on the $m_\chi$ dependence of the higher order corrections.

Despite the discrepancy between our result and that of HI, we emphasise that we have confidence in our result as stated. This confidence is derived from the non-trivial cross checks we have performed on our result. In detail, these are
\begin{itemize}
\item The cancellation in the $\mathcal{O}(\alpha_2)$ corrections of the $\mu_{m_{\chi}}$ dependence in our high-scale matching coefficients, stated in Eq.~\eqref{eq:WilsonCoeff}, with the high-scale dependence entering from the anomalous dimension, as stated in Eqs.~\eqref{eq:anomdim} and \eqref{eq:anomdimparts}. This cancellation is demonstrated in App.~\ref{app:consistency};
\item In the absence of running, the cancellation in the $\mathcal{O}(\alpha_2)$ corrections of the $\mu$ dependence between our high and low-scale results, where the latter is stated in Eqs.~\eqref{eq:lowbreakdown}, \eqref{eq:LowSoft}, \eqref{eq:LowColinear}, \eqref{eq:LowColinearConsts1}, and \eqref{eq:LowColinearConsts2}. This cancellation also depends on the SM SU(2)$_{\rm L}$ and U(1)$_Y$ $\beta$-functions and is shown in App.~\ref{app:consistencylow};
\item We have confirmed that the $\mu$ dependence in our low-scale result matches that in \cite{Bauer:2014ula}, when we eliminate parts of our calculation in order to make the same assumptions used in that work;
\item The form of the dominant $\mu$ independent terms in the low-scale matching are in agreement with the results of \cite{Chiu:2007yn,Chiu:2007dg,Chiu:2008vv,Chiu:2009mg,Chiu:2009ft}, as discussed in App.~\ref{app:lowscalematching}; and
\item We have confirmed that the framework used to calculate the low-scale matching for our non-relativistic initial state kinematics, reproduces the results of \cite{Chiu:2007yn,Chiu:2007dg,Chiu:2008vv,Chiu:2009mg,Chiu:2009ft} when we instead consider massless initial states as used in those references.
\end{itemize}

\section{Conclusion}
\label{sec:conclusion}

In this work we provide analytic expressions for the full one-loop corrections to heavy wino dark matter annihilation, allowing the systematic resummation of electroweak Sudakov logarithms to NLL$^{\prime}$ for the line cross section. We have compared our result to earlier numerical calculations of such effects, finding results similar in behaviour but quantitatively different. Our result is stated in a manner that can be straightforwardly extended to higher order, with our result already reducing the perturbative uncertainty from Sudakov effects on this process to $\mathcal{O}(1\%)$.

\newpage
\section*{Acknowledgements}

The authors thank Aneesh Manohar for helpful discussions and for providing explicit cross checks for aspects of our low-scale matching. We also thank Matthew Baumgart, Tim Cohen, Ian Moult and Hiren Patel for helpful discussions and comments. Feynman diagrams were drawn using \cite{Ellis:2016jkw} and NLR thanks Joshua Ellis for assistance with its use. This work is supported by the U.S. Department of Energy under grant Contract Numbers DE-SC00012567, DE-SC0013999, DE-SC0011090 and by the Simons Foundation Investigator grant 327942. NLR is supported in part by the American Australian Association’s ConocoPhillips Fellowship.

\clearpage
\appendix
\section{One-loop Calculation of $\chi^a \chi^b \to W^c W^d$ in the Full Theory}
\label{app:oneloopfull}

In this appendix we outline the details of the high-scale matching calculation, which gives rise to the Wilson coefficients stated in Eq.~\eqref{eq:WilsonCoeff}. These coefficients are determined solely by the ultraviolet (UV) physics, allowing us to simplify the calculation by working in the unbroken theory with $m_W=m_Z=\delta m=0$. Combining this with the heavy Majorana fermion DM being non-relativistic, there are only two possible Dirac structures that can appear in the result:
\begin{eqnarray}
\mathcal{M}_A &=& \epsilon_{\mu}^*(p_3) \epsilon_{\nu}^*(p_4) \epsilon^{\sigma \mu \nu \alpha} p_{3\alpha} i \bar{v}(p_2) \gamma_{\sigma} \gamma_5 u(p_1)\,, \nn
\mathcal{M}_B &=& \epsilon_{\mu}^*(p_3) \epsilon_{\nu}^*(p_4) g^{\mu \nu} \bar{v}(p_2) \slashed{p}_3 u(p_1)\,,
\label{eq:FullTheoryOps}
\end{eqnarray}
where $p_1$ and $p_2$ are the momenta of the incoming fermions, whilst $p_3$ and $p_4$ correspond to the outgoing bosons. The symmetry properties of these structures under the interchange of initial and final state particles allow us to write our full amplitude as:
\begin{eqnarray}
\mathcal{M}_{abcd} &=& \frac{4\pi\alpha_2}{m_{\chi}^2} \left\{ \left[ B_1 \delta_{ab} \delta_{cd} + B_2 \left( \delta_{ac} \delta_{bd} + \delta_{ad} \delta_{bc} \right) \right] \mathcal{M}_A \right.\nn
&&\left.+ B_3 \left( \delta_{ac} \delta_{bd} - \delta_{ad} \delta_{bc} \right) \mathcal{M}_B \right\}\,.
\label{eq:FullTheoryStructure}
\end{eqnarray}
The above equation serves to define the Wilson coefficients $B_r$ in a convenient form. These coefficients are related to the EFT coefficients of the operators defined in Eq.~\eqref{eq:Ops} and \eqref{eq:GaugeIndex} via:
\begin{equation}\begin{aligned}
C_1 = (- \pi \alpha_2/m_{\chi}) B_1\,,\;\;\;\;C_2 = (-2\pi \alpha_2/m_{\chi}) B_2\,.
\label{eq:CoeffMapping}
\end{aligned}\end{equation}
For NLL accuracy we only need the tree-level value of these coefficients, which receive a contribution from $s$, $t$ and $u$-channel type graphs and were calculated in \cite{Ovanesyan:2014fwa}. For completeness we state their values here:
\begin{equation}\begin{aligned}
B_1^{(0)} = 1\,,\;\;\; B_2^{(0)} = - \frac{1}{2}\,,\;\;\; B_3^{(0)} = 0\,.
\label{eq:TreeLevelCoeff}
\end{aligned}\end{equation}
Combining these with Eq.~\eqref{eq:CoeffMapping}, we see that the first terms in Eq.~\eqref{eq:WilsonCoeff} are indeed the tree-level contributions as claimed.

The operator associated with $B_3$ was not discussed in the earlier work of \cite{Ovanesyan:2014fwa} as it cannot contribute to the high-scale matching calculation at any order, as we will now argue. Firstly note that the $B_3$ operator is skew under the interchange $a \leftrightarrow b$. Due to the mass splitting between the neutral and charged states, present day annihilation is initiated purely by $\chi^0 \chi^0 = \chi^3 \chi^3$, a symmetric state that cannot overlap with $B_3$. One may worry that exchange of one or more weak bosons between the initial states -- the hallmark of the Sommerfeld enhancement -- may nullify this argument. But it can be checked that if the initial states to such an exchange have identical gauge indices, then so will the final states. As such $B_3$ is not relevant for calculating high-scale matching.\footnote{Diagrams where a soft gauge boson is exchanged between an initial and final state particle would in principle allow $B_3$ to contribute. Such a contributions would however be to the low-scale matching, which we discuss in App.~\ref{app:lowscalematching}. As discussed there, $B_3$ contributions to present day DM annihilation are power suppressed, and therefore do not contribute at any order in the leading power effective theory.}

In spite of this there are several reasons to calculate $B_3$ here. From a practical point of view $B_3$ gives us an additional handle on the consistency of our result, which we check in App.~\ref{app:consistency}. Given that many graphs that generate $B_1$ and $B_2$ also contribute to $B_3$, the consistency of $B_3$ provides greater confidence in the results for the operators we are interested in. Further, from a physics point of view, although $B_3$ is not relevant for high-scale matching when considering present day indirect detection experiments, it could be relevant for calculating the annihilation rate in the early universe, where all states in the DM triplet were present, to the extent that the non-relativistic approximation is still relevant. For this reasons we state it in case it is of interest for future work, such as expanding on calculations of the relic density at one loop (see for example \cite{Boudjema:2005hb,Baro:2007em,Baro:2009na}).

\subsection*{Determining Matching Coefficients}

Let us briefly review how matching coefficients are calculated at one loop. To begin with we can write the general structure of the UV and infrared (IR) divergences of the bare one-loop result for annihilation diagrams in the full theory as:
\begin{equation}
\mathcal{M}_{\rm bare}^{\rm full} = \frac{K}{\epsilon_{\rm IR}^2} + \frac{L}{\epsilon_{\rm IR}} + \frac{M}{\epsilon_{\rm UV}} + N \left( \frac{1}{\epsilon_{\rm UV}} - \frac{1}{\epsilon_{\rm IR}} \right) + C\,,
\label{eq:barefull}
\end{equation}
where $N$ is the coefficient associated with the various scaleless integrals, and $C$ is the finite contribution. Now the full theory is a renormalizable gauge theory, so we know the additional counter-term and wavefunction renormalization contributions must be of the form:
\begin{equation}
\delta^{\rm full} = - \frac{M+N}{\epsilon_{\rm UV}} + D + \frac{E}{\epsilon_{\rm IR}^2} + \frac{F}{\epsilon_{\rm IR}}\,,
\label{eq:fullct}
\end{equation}
where the values of $D$, $E$ and $F$ are scheme dependent. Nonetheless when calculating matching coefficients it is easiest to work in the on-shell scheme for the wave-function renormalization factors, so below to denote this we add an ``os'' subscript to $D$, $E$ and $F$. The reason this scheme is the most straightforward, is that in any other scheme when we map our Feynman amplitude calculation for $\mathcal{M}^{\rm full}$ onto the $S$-matrix elements we want via the LSZ reduction, there will be non-trivial residues corresponding to the external particles. When using the on-shell scheme for the wave-function renormalization factors, however, these residues are just unity, which simplifies the calculation as we can then ignore them. We emphasise that whatever scheme one uses, the final result for the Wilson coefficients in $\overline{\rm MS}$ will be the same.

With this in mind, if we then combine $\delta^{\rm full}$ with the bare results we obtain a UV finite answer:
\begin{equation}
\mathcal{M}_{\rm ren.}^{\rm full} = \frac{K+E_{\rm os}}{\epsilon_{\rm IR}^2} + \frac{L-N+F_{\rm os}}{\epsilon_{\rm IR}} + C + D_{\rm os}\,.
\label{eq:renfull}
\end{equation}
In our calculation we will use dimensional regularization to regulate both UV and IR divergences, which effectively sets $\epsilon_{\rm UV} = \epsilon_{\rm IR}$, causing all scaleless integrals to vanish. Naively this seems to change the above argument, but as long as we still use the correct counter-term in Eq.~\eqref{eq:fullct} we find:
\begin{equation}\begin{aligned}
\mathcal{M}_{\rm ren.}^{\rm full} =& \frac{K}{\epsilon^2} + \frac{L}{\epsilon} + \frac{M}{\epsilon} + C - \frac{M+N}{\epsilon} \\
&+ D_{\rm os} + \frac{E_{\rm os}}{\epsilon^2} + \frac{F_{\rm os}}{\epsilon} \\
=& \frac{K+E_{\rm os}}{\epsilon^2} + \frac{L-N+F_{\rm os}}{\epsilon} + C + D_{\rm os}\,.
\end{aligned}\end{equation}
Comparing this with Eq.~\eqref{eq:renfull}, we see that if we interpret all of the divergences in the final result as IR, then this method is equivalent to carefully distinguishing $\epsilon_{\rm UV}$ and $\epsilon_{\rm IR}$ throughout.

In the EFT, with the above choice of zero masses and working on-shell with dimensional regularization, all graphs are scaleless. At one loop they have the general form:\footnote{One may worry there could also be scaleless integrals of the form $\left( \epsilon_{\rm UV}^{-1} - \epsilon_{\rm IR}^{-1} \right)^2$, but the use of the zero-bin subtraction \cite{Manohar:2006nz} ensures such contributions cannot appear.}
\begin{equation}
\mathcal{M}_{\rm bare}^{\rm EFT} = O \left( \frac{1}{\epsilon_{\rm UV}^2} - \frac{1}{\epsilon_{\rm IR}^2} \right) + P \left( \frac{1}{\epsilon_{\rm UV}} - \frac{1}{\epsilon_{\rm IR}} \right)\,.
\label{eq:bareeft}
\end{equation}
Importantly if we have the correct EFT description of the full theory, then the two theories must have the same IR divergences. Comparing Eq.~\eqref{eq:bareeft} to Eq.~\eqref{eq:renfull}, we see this requires $O = - K-E_{\rm os}$ and $P=N-L-F_{\rm os}$. The EFT is again a renormalizable theory, so we can cancel the UV divergences using $\delta^{\rm EFT} = (K+E_{\rm os}) \epsilon_{\rm UV}^{-2} +(L+F_{\rm os}-N) \epsilon_{\rm UV}^{-1}$. Note as all EFT graphs are scaleless there are no finite contributions that could be absorbed into the counter-term, so in any scheme there is no finite correction to $\delta^{\rm EFT}$. Using this counter-term, we conclude:
\begin{equation}
\mathcal{M}_{\rm ren.}^{\rm EFT} = \frac{K+E_{\rm os}}{\epsilon_{\rm IR}^2} + \frac{L-N+F_{\rm os}}{\epsilon_{\rm IR}} \,.
\label{eq:reneft}
\end{equation}
Again note that for a similar argument to that in the full theory, if we had set $\epsilon_{\rm UV} = \epsilon_{\rm IR}$ at the outset, then as long as we still used the correct counter-term we would arrive at the same result.

The matching coefficient is then obtained from subtracting the renormalized EFT from the renormalized full theory result, so taking the appropriate results above we conclude:
\begin{equation}
\mathcal{M}_{\rm ren.}^{\rm full} - \mathcal{M}_{\rm ren.}^{\rm EFT} = C+D_{\rm os}\,.
\label{eq:sketchmatching}
\end{equation}
Comparing this with Eq.~\eqref{eq:renfull}, we see that provided we have the correct EFT, then the matching coefficient is just the finite contribution to the renormalized full-theory amplitude in the on-shell scheme. Even though this result makes explicit reference to a scheme in $D_{\rm on-shell}$, it is in fact scheme independent. The reason for this is that if we worked in a different scheme, although $D$ would change, we would also have to account for the now non-trivial external particle residues that enter via LSZ. Their contribution is what ensures Eq.~\eqref{eq:sketchmatching} is scheme independent.

\subsection*{Results of the Calculation}

As outlined above, in order to obtain the matching coefficients we need the finite contribution to the renormalized full theory amplitude. Now to compute this in the particular theory we consider in this paper, we need to calculate the 25 diagrams that contribute to the one-loop correction to $\chi^a \chi^b \to W^c W^d$. The diagrams are identical to those considered in \cite{Fuhrer:2010eu}, where they defined a numbering scheme for the diagrams, grouping them by topology and labelling them as $T_i$ for various $i$. We follow that numbering scheme here, but cannot use their results as they considered massless initial state fermions whilst ours are massive and non-relativistic. In general we calculate the diagrams using dimensional regularization with $d=4-2\epsilon$ to regulate the UV and IR, and work in `t Hooft-Feynman gauge. Loop integrals are determined using Passarino-Veltman reduction \cite{Passarino:1978jh}, and we further make use of the results in \cite{Denner:1992vza,Ellis:2007qk,Hooft:1978xw,Ellis:2011cr} as well as FeynCalc \cite{Mertig:1990an,Shtabovenko:2016sxi} and Package-X \cite{Patel:2015tea}.

In the EFT description of the full theory outlined in Sec.~\ref{sec:NLL}, the factorization of the matrix elements ensured a separation between the Sommerfeld and Sudakov contributions. Yet for the full theory no clear separation exists and there will be graphs that contribute to both effects -- in particular the graph $T_{1c}$ considered below. The purpose of the Wilson coefficients we are calculating here is to provide corrections to the Sudakov contribution -- we do not want to spoil the EFT distinction by including Sommerfeld effects in these coefficients. In order to cleanly separate the contributions we take the relative velocity of our non-relativistic initial states to be zero. This ensures that any contribution of the form $1/v$, characteristic of Sommerfeld enhancement, become power divergences and therefore vanish in dimensional regularization. This is different to the treatment in HI, where they calculated the diagram without sending $v \to 0$ and subtracted the Sommerfeld contribution by hand. Our treatment is known from studies in NRQCD~\cite{Beneke:1997zp,Luke:1999kz,Brambilla:2004jw,Manohar:2000hj} (for example) to give the same result as calculating at finite $v$ and subtracting the NRDM-SCET$_{\rm EW}$ Sommerfeld graphs. 

In our calculation the DM is a Majorana fermion. It turns out that for almost all the graphs below the result is identical regardless of whether we think of the fermion as Majorana or Dirac -- a result that is also true at tree-level. The additional symmetry factors in the Majorana case are exactly cancelled by the factors of $1/2$ entering from the Majorana Lagrangian. The exceptions to this are for graphs containing a closed loop of fermions, specifically $T_{2d}$ and $T_{6d}$ below, as well as closed fermion loop contributions to the counter-terms.

Using the approach outlined above we now state the contribution to $B_r$ as defined in Eq.~\eqref{eq:FullTheoryStructure} graph by graph. Throughout we define $L\equiv \ln \mu/2m_{\chi}$.

\subsection*{\large $T_{1a}$}
\begin{center}
\includegraphics[height=0.15\columnwidth]{./Plots/{T1a}.pdf}
\end{center}
The result for this graph and its cross term is:
\begin{equation}\begin{aligned}
B_1^{[1a]} &= \frac{\alpha_2}{4\pi} \left[ - \frac{2}{\epsilon^2} - \frac{1}{\epsilon} \left( 4L + 2 i \pi + 2 \right) - 4 L^2 \right. \\
&\hspace{0.4in} \left.- 4 L - 4 i \pi L - 4 + \frac{7\pi^2}{6} + 4 \ln 2  \right]\,, \\
B_2^{[1a]} &= \frac{1}{2} B_1^{[1a]}\,, \\
B_3^{[1a]} &= \frac{\alpha_2}{4\pi} \left[ \frac{1}{4 \epsilon^2} + \frac{1}{4 \epsilon} \left( 2L - 3 i \pi - 2 \right) + \frac{1}{2} L^2 \right. \\
&\hspace{1.04in}- L - \frac{3}{2} i \pi L + \frac{17\pi^2}{48}\\
&\hspace{1.04in}\left. - \frac{1}{6} (2 + 7 i \pi - 8 \ln 2 ) \right]\,.
\end{aligned}\end{equation}
In calculating this graph in the non-relativistic limit via Passarino-Veltman reduction there are additional spurious divergences that must be regulated. The origin of these divergences is that Passarino-Veltman assumes the momenta appearing in the integrals to be linearly independent. But in the center of momentum frame if we take $v=0$, then $p_1$ and $p_2$ are identical and this assumption breaks down, leading to the divergences of the form $(s-4m_{\chi}^2)^{-1}$, where $s=(p_1+p_2)^2$. A simple way to regulate them is to give the initial states a small relative velocity. This does not lead to a violation of our separation of Sommerfeld and Sudakov effects as this graph does not contribute to the Sommerfeld enhancement. As such this procedure introduces no $1/v$ contributions to the final result and the regulator can be safely removed at the end. This is the only diagram where this issue appears -- if it occurred in a graph that did contribute to the Sommerfeld effect we would need to use a different regulator, or explicitly subtract the corresponding EFT graph at finite $v$. 

\subsection*{\large $T_{1b}$}
\begin{center}
\includegraphics[height=0.15\columnwidth]{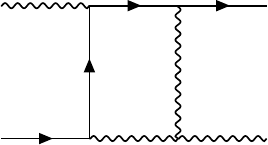}
\end{center}
This graph has a single crossed term and combining the two yields:
\begin{equation}\begin{aligned}
B_1^{[1b]} &= B_3^{[1b]} = 0\,, \\
B_2^{[1b]} &= \frac{\alpha_2}{4\pi} \left[ \frac{2}{\epsilon^2} + \frac{4L+2}{\epsilon} + 4L(L+1) \right. \\
&\hspace{0.85in} \left. - \frac{2\pi^2}{3} + 4 - 8 \ln 2 \right]\,.
\end{aligned}\end{equation}

\subsection*{\large $T_{1c}$}
\begin{center}
\includegraphics[height=0.15\columnwidth]{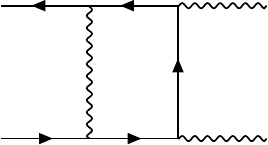}
\end{center}
The combination of this graph and its crossed term is:
\begin{equation}\begin{aligned}
B_1^{[1c]} &= \frac{\alpha_2}{4\pi} \left[ \frac{2}{\epsilon} - 4 + 4L + 4 \ln 2 \right]\,, \\
B_2^{[1c]} &= \frac{1}{2} B_1^{[1c]}\,, \\
B_3^{[1c]} &= \frac{\alpha_2}{4\pi} \left[ \frac{1}{\epsilon} - 2 + 2L + \frac{\pi^2}{4} - 2 \ln 2 \right]\,.
\end{aligned}\end{equation}
Formally this graph also gives a contribution to the Sommerfeld enhancement in the full theory. Nevertheless as we take $v=0$ at the outset, the contribution here is purely to the Sudakov terms.

\subsection*{\large $T_{1d}$}
\begin{center}
\includegraphics[height=0.15\columnwidth]{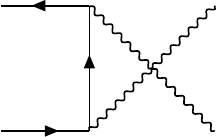}
\end{center}
The contribution from this diagram vanishes in the non-relativistic limit, i.e.
\begin{equation}\begin{aligned}
B_1^{[1d]} = B_2^{[1d]} = B_3^{[1d]} = 0\,.
\end{aligned}\end{equation}

\subsection*{\large $T_{2a}$}
\begin{center}
\includegraphics[height=0.15\columnwidth]{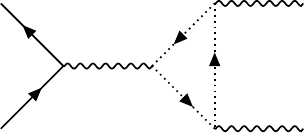}
\end{center}
For the case of ghosts running in the loop of the above graph we have its contribution and the crossed term giving
\begin{equation}\begin{aligned}
B_1^{[2a]} &= B_2^{[2a]} = 0\,, \\
B_3^{[2a]} &= \frac{\alpha_2}{4\pi} \left[ \frac{1}{24\epsilon} + \frac{2L + i \pi}{24} + \frac{11}{72} \right]\,. 
\end{aligned}\end{equation}

\subsection*{\large $T_{2b}$}
\begin{center}
\includegraphics[height=0.15\columnwidth]{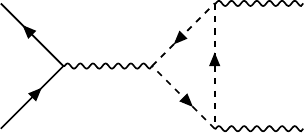}
\end{center}
For a scalar Higgs in the loop, the graph and its cross term contribute:
\begin{equation}\begin{aligned}
B_1^{[2b]} &= B_2^{[2b]} = 0\,, \\
B_3^{[2b]} &= \frac{\alpha_2}{4\pi} \left[ \frac{1}{12\epsilon} + \frac{2L + i \pi}{12} + \frac{11}{36} \right]\,.
\end{aligned}\end{equation}

\subsection*{\large $T_{2c}$}
\begin{center}
\includegraphics[height=0.15\columnwidth]{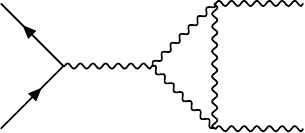}
\end{center}
There is no crossed graph associated with the graph above as the gauge bosons running in the loop are real fields. As such taking just this graph gives
\begin{equation}\begin{aligned}
B_1^{[2c]} &= B_2^{[2c]} = 0\,, \\
B_3^{[2c]} &= \frac{\alpha_2}{4\pi} \left[ \frac{3}{4\epsilon^2} + \frac{1}{\epsilon} \left( \frac{3}{4} (2L + i \pi) + \frac{17}{8} \right) \right.\\
&\hspace{0.5in}+ \frac{3}{8} \left( 2L + i \pi \right)^2 \\
&\hspace{0.5in}\left. + \frac{17}{8} \left( 2L + i \pi \right) + \frac{95}{24} - \frac{\pi^2}{16} \right]\,.
\end{aligned}\end{equation}

\subsection*{\large $T_{2d}$}
\begin{center}
\includegraphics[height=0.15\columnwidth]{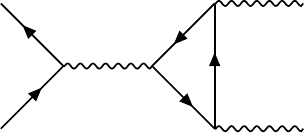}
\end{center}
There are two types of fermions that can run in the loop: the Majorana triplet fermion that make up our DM or left-handed SM doublets. As with the gauge bosons these SM fermions are taken to be massless and for generality we say there are $n_D$ of them.\footnote{For the SM well above the electroweak scale $n_D=12$. In detail, for each generation there are four doublets: the lepton doublet and due to color, three quark doublets. As such for three generations we have twelve left-handed SM doublets.} For the SM doublets there is a crossed graph, whilst for the Majorana DM field there is not, so that:
\begin{equation}\begin{aligned}
B_1^{[2d]} &= B_2^{[2d]} = 0\,, \\
B_3^{[2d]} &= \frac{\alpha_2}{4\pi} \left[ - \left( \frac{2}{3\epsilon} + \frac{4}{3} L + \frac{4}{3} \ln 2 - \frac{5}{9} + \frac{\pi^2}{4} \right) \right. \\
&\hspace{0.5in} \left.- n_D \left( \frac{1}{6\epsilon} + \frac{1}{6} \left( 2L + i \pi \right) + \frac{7}{36} \right) \right]\,.
\end{aligned}\end{equation}
If the DM had been a Dirac field instead, there would have been a crossed graph and the result would be modified such that the first line of $B_3^{[2d]}$ gets multiplied by 2.

The factor of $7/36$ we find in the last line of $B_3^{[2d]}$ is consistent with the expression found for this graph, but with different kinematics, in \cite{Fuhrer:2010eu}, but disagrees with \cite{Bardin:1999ak}.

\subsection*{\large $T_{2e-h}$}
\begin{center}
\includegraphics[height=0.15\columnwidth]{./Plots/{T2e}.pdf} \hspace{0.1in}
\includegraphics[height=0.15\columnwidth]{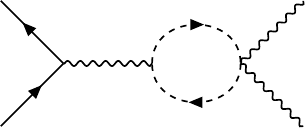} \\
\vspace{0.1in}
\includegraphics[height=0.15\columnwidth]{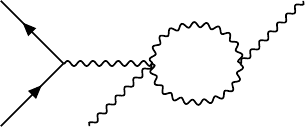} \hspace{0.1in}
\includegraphics[height=0.15\columnwidth]{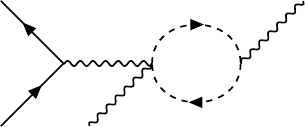}
\end{center}
The four graphs shown above do not contribute to our one-loop result; the graphs on the top row vanish at leading order for non-relativistic initial states, whilst the loops on the second line are both scaleless and so are identically zero in dimensional regularization. As such we have:
\begin{equation}\begin{aligned}
B_1^{[2e-f]} = B_2^{[2e-f]} = B_3^{[2e-f]} = 0\,.
\end{aligned}\end{equation}

\subsection*{\large $T_{3a}$ and $T_{4a}$}
\begin{center}
\includegraphics[height=0.15\columnwidth]{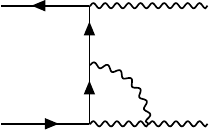} \hspace{0.1in}
\includegraphics[height=0.15\columnwidth]{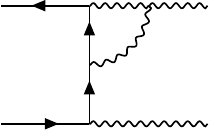}
\end{center}
The two graphs shown above have identical amplitudes. For each graph independently, the sum of it and its crossed graph is:
\begin{equation}\begin{aligned}
B_1^{[3a/4a]} &= \frac{\alpha_2}{4\pi} \left[ - \frac{1}{\epsilon^2} + \frac{2-2L}{\epsilon} - 2L^2 \right. \\
&\hspace{0.3in} \left. + 4L - 2\ln 2 + 4 + \frac{\pi^2}{12} \right]\,, \\
B_2^{[3a/4a]} &= - \frac{1}{2} B_1^{[3a/4a]}\,, \\
B_3^{[3a/4a]} &= \frac{1}{2} B_1^{[3a/4a]}\,.
\end{aligned}\end{equation}

\subsection*{\large $T_{3b}$ and $T_{4b}$}
\begin{center}
\includegraphics[height=0.15\columnwidth]{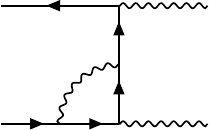} \hspace{0.1in}
\includegraphics[height=0.15\columnwidth]{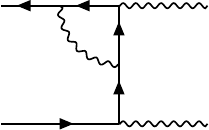}
\end{center}
As for $T_{3a}$ and $T_{4a}$, these two graphs also have equal amplitudes. Again we provide the combination of each with its crossed graph:
\begin{equation}\begin{aligned}
B_1^{[3b/4b]} &= \frac{\alpha_2}{4\pi} \left[ \frac{1}{\epsilon} + 2L - 2\ln 2 + \frac{\pi^2}{4} \right]\,, \\
B_2^{[3b/4b]} &= -\frac{1}{2} B_1^{[3b/4b]}\,, \\
B_3^{[3b/4b]} &= \frac{1}{2} B_1^{[3b/4b]}\,.
\end{aligned}\end{equation}

\subsection*{\large $T_{5a}$}
\begin{center}
\includegraphics[height=0.15\columnwidth]{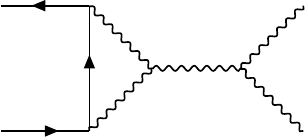}
\end{center}
Whether the above graph has a crossed graph associated with interchanging the initial states depends on the identity of the initial state fermions. For Majorana fermions there is such a crossing, whilst for Dirac there is not. Despite this, in either case the combination of the graph and its crossing (where it exists) is the same in both cases and is simply:
\begin{equation}\begin{aligned}
B_1^{[5a]} &= B_2^{[5a]} = 0\,, \\ 
B_3^{[5a]} &= \frac{\alpha_2}{4\pi} \left[ - \frac{3}{2\epsilon} - 3 L - \frac{13}{3} \ln 2 - \frac{8}{3} + \frac{2}{3} i \pi \right]\,.
\end{aligned}\end{equation}

\subsection*{\large $T_{5b}$}
\begin{center}
\includegraphics[height=0.15\columnwidth]{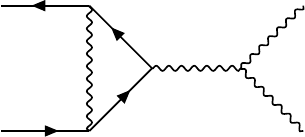}
\end{center}
As for $T_{5a}$ the existence of a crossed graph depends on the nature of the DM. Regardless again the result is the same if we take it to be Dirac or Majorana, which is:
\begin{equation}\begin{aligned}
B_1^{[5b]} &= B_2^{[5b]} = 0\,, \\ 
B_3^{[5b]} &= \frac{\alpha_2}{4\pi} \left[ \frac{3}{2\epsilon} + 3L + 3 \ln 2 - 2 \right]\,.
\end{aligned}\end{equation}

\subsection*{\large $T_{6a}$}
\begin{center}
\includegraphics[height=0.15\columnwidth]{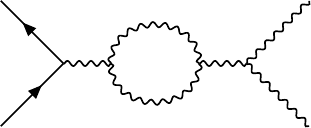}
\end{center}
For a gauge boson in the loop we have:
\begin{equation}\begin{aligned}
B_1^{[6a]} &= B_2^{[6a]} = 0\,, \\ 
B_3^{[6a]} &= \frac{\alpha_2}{4\pi} \left[ - \frac{19}{12\epsilon} - \frac{19}{6} L - \frac{29}{9} - \frac{19}{12} i \pi \right]\,.
\end{aligned}\end{equation}
Note this graph and the remaining $T_6$ type topologies have no crossed graphs.

\subsection*{\large $T_{6b}$}
\begin{center}
\includegraphics[height=0.15\columnwidth]{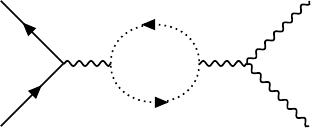}
\end{center}
In the case of a ghost loop we have:
\begin{equation}\begin{aligned}
B_1^{[6b]} &= B_2^{[6b]} = 0\,, \\ 
B_3^{[6b]} &= \frac{\alpha_2}{4\pi} \left[ - \frac{1}{12\epsilon} - \frac{1}{6} L - \frac{2}{9} - \frac{1}{12} i \pi \right]\,.
\end{aligned}\end{equation}

\subsection*{\large $T_{6c}$}
\begin{center}
\includegraphics[height=0.15\columnwidth]{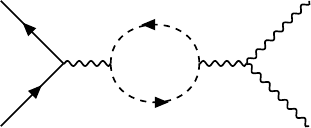}
\end{center}
For a scalar Higgs we have an identical contribution to $T_{6b}$:
\begin{equation}\begin{aligned}
B_1^{[6c]} &= B_2^{[6c]} = 0\,, \\ 
B_3^{[6c]} &= \frac{\alpha_2}{4\pi} \left[ - \frac{1}{12\epsilon} - \frac{1}{6} L - \frac{2}{9} - \frac{1}{12} i \pi \right]\,.
\end{aligned}\end{equation}

\subsection*{\large $T_{6d}$}
\begin{center}
\includegraphics[height=0.15\columnwidth]{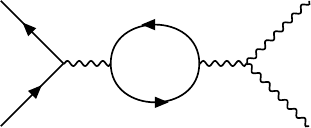}
\end{center}
As for $T_{2d}$ the fermion in the loop could again be either DM or SM. Allowing there to be $n_D$ left-handed SM doublets we have
\begin{equation}\begin{aligned}
B_1^{[6d]} &= B_2^{[6d]} = 0\,, \\ 
B_3^{[6d]} &= \frac{\alpha_2}{4\pi} \left[ \left( \frac{2}{3\epsilon} + \frac{4}{3} L + \frac{4}{3} \ln 2 + \frac{16}{9} \right) \right. \\
&\hspace{0.5in} \left.+ n_D \left( \frac{1}{6\epsilon} + \frac{1}{3} L + \frac{5}{18} + \frac{1}{6} i \pi \right) \right]\,.
\end{aligned}\end{equation}
Here there is a symmetry factor of $1/2$ for the loop in the case of the Majorana DM field. If the DM was a Dirac fermion instead, the first line of $B_3^{[6d]}$ would get multiplied by $2$ as this symmetry factor would not be present. Note that whilst the divergence would change in the Dirac case, this is compensated by the associated modification to $T_{2d}$.

\subsection*{\large $T_{6e}$ and $T_{6f}$}
\begin{center}
\includegraphics[height=0.15\columnwidth]{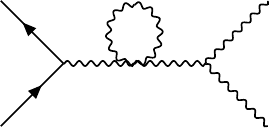} \hspace{0.1in}
\includegraphics[height=0.15\columnwidth]{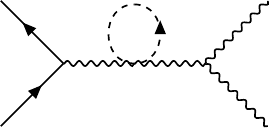}
\end{center}
Both of these integrals are scaleless and vanish in dimensional regularization, so:
\begin{equation}\begin{aligned}
B_1^{[6e-f]} = B_2^{[6e-f]} = B_3^{[6e-f]} = 0\,.
\end{aligned}\end{equation}

\subsection*{\large $T_{7}$}
\begin{center}
\includegraphics[height=0.18\columnwidth]{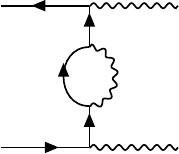}
\end{center}
For the final graph we again have a crossed contribution, and combining the two gives:
\begin{equation}\begin{aligned}
B_1^{[7]} &= \frac{\alpha_2}{4\pi} \left[ -\frac{8}{\epsilon} - 16L - 12 \right]\,, \\
B_2^{[7]} &= - \frac{1}{2} B_1^{[7]}\,, \\
B_3^{[7]} &= \frac{1}{2} B_1^{[7]}\,.
\end{aligned}\end{equation}

\subsection*{Counter-terms}
To begin with, as $B_3$ vanishes at tree level there are no counter-term corrections to its value at one loop. Instead we only need to consider graphs that would contribute to $B_1$ and $B_2$, of which there are three:
\begin{center}
\includegraphics[height=0.15\columnwidth]{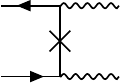} \hspace{0.1in}
\includegraphics[height=0.16\columnwidth]{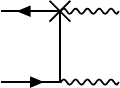} \hspace{0.1in}
\includegraphics[height=0.16\columnwidth]{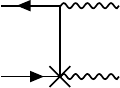}
\end{center}
The graph on the left corresponds to the internal wave-function and mass renormalization of the DM -- renormalization factors denoted as $Z_{\chi}$ and $Z_m$ -- whilst the remaining two graphs account for the renormalization of the DM and electroweak gauge boson interaction vertex $g_2 \bar{\chi} \slashed{W} \chi$ -- here $Z_1$ (which includes coupling and external line wavefunction renormalization). Now if we calculate the above three graphs, we find a contribution proportional to the tree-level amplitude $\mathcal{M}_{\rm tree}$, as well as a term that would contribute to $B_3$. The contribution to $B_3$ is cancelled by the additional $s$-channel type counter-term graphs not drawn, so the full counter-term contribution leaves only:
\begin{equation}
\left(2 \delta_1 - \delta_{\chi} - \delta_m \right) \mathcal{M}_{\rm tree}\,,
\label{eq:CTgraph}
\end{equation}
where we have used $Z_i = 1 + \delta_i$.

Next, when determining the $\delta_i$ we need to pick a scheme. As explained above, when calculating matching coefficients it is easiest to work in the on-shell scheme for wavefunction renormalization to ensure we do not have to worry about residues from the LSZ reduction. The meaning of the on-shell values of $\delta_{\chi}$ and $\delta_m$ is clear, whereas for $\delta_1$ we must write this out more explicitly. By definition we know $\delta_1 = \delta_{g_2} + \frac{1}{2} \delta_W + \delta_{\chi}$, where $\delta_{g_2}$ and $\delta_W$ are the counter-terms for the coupling and gauge boson wave-functions respectively. For the gauge-boson wave-function we use the on-shell scheme as usual. For the coupling counter-term, however, we define it in the $\overline{\rm MS}$ scheme. Since our full theory is defined with the DM as a propagating degree of freedom, this coupling is defined above the $m_{\chi}$. In the EFT the DM is integrated out, so the appropriate coupling for the matching is one defined below $m_{\chi}$. We put this issue aside for now and return to it in the next section.

The above choices then define our scheme for $\delta_1$ in a manner that ensures all residues are still 1. With this scheme, we can then calculate the relevant counter-terms for Majorana DM and find:\footnote{As pointed out in~\cite{Beneke:2018ssm}, the original version of this paper had incorrect coefficients in some entries of these counter terms. This affected the constant terms appearing in $C_1$ and $C_2$ in Eq.~(\ref{eq:WilsonCoeff}), namely the $+12$ and $+2$ terms respectively. The resulting changes to the figures are invisible to the eye. We thank the authors of that work for bringing this to our attention.}
\begin{align} \label{eq:CTvalues}
\delta_{\chi} &= - \frac{\alpha_2}{4\pi} \left[ \frac{2}{\epsilon_{\rm UV}} + \frac{4}{\epsilon_{\rm IR}} +  12L + 12 \ln 2 + 8 \right]\,,  \\
\delta_m &= - \frac{\alpha_2}{4\pi} \left[ \frac{6}{\epsilon_{\rm UV}} + 12L + 12 \ln 2 + 8 \right]\,, \nonumber
\end{align}
\begin{align}
\delta_W &= - \frac{\alpha_2}{4\pi} \left[ \frac{2n_D-11}{6\epsilon_{\rm UV}} + \frac{19-2n_D}{6\epsilon_{\rm IR}} + \frac{8}{3} L + \frac{8}{3} \ln 2 \right]\,, \nonumber \\
\delta_{g_2} &= - \frac{\alpha_2}{4\pi} \left[ \frac{35-2n_D}{12\epsilon_{\rm UV}} \right]\,,\nonumber \\
\delta_1 &= - \frac{\alpha_2}{4\pi} \left[ \frac{4}{\epsilon_{\rm UV}} + \frac{67-2n_D}{12\epsilon_{\rm IR}} + \frac{40}{3} L + \frac{40}{3} \ln 2 + 8 \right]\,, \nonumber
\end{align}
where $n_D$ is again the number of left-handed SM doublets. Recall that in determining the counter-terms we cannot neglect scaleless integrals as we did for the main calculation, so their contribution has been included here and we explicitly distinguish $\epsilon_{\rm UV}$ from $\epsilon_{\rm IR}$. Substituting these results into Eq.~\eqref{eq:CTgraph}, we find the crossed contribution is:
\begin{equation}\begin{aligned}
B_1^{[{\rm CT}]} &= \frac{\alpha_2}{4\pi} \left[ \frac{2n_D-43}{6\epsilon_{\rm IR}} - \frac{8}{3} L - \frac{8}{3} \ln 2  \right]\,, \\
B_2^{[{\rm CT}]} &= - \frac{1}{2} B_1^{[{\rm CT}]}\,, \\
B_3^{[{\rm CT}]} &= 0\,.
\label{eq:CTcontribution}
\end{aligned}\end{equation}
Interestingly the counter-term contribution is UV finite. This implies that the sum of all one-loop graphs before adding in counter-terms must be UV finite. Given that we used dimensional regularization to regulate both UV and IR divergences this cannot be immediately read off from our results, but going back to the integrals and keeping track of the UV divergences we confirmed that the sum is indeed UV finite. 

Note if our DM field had instead been a Dirac fermion, the above result would vary, but in a way that is exactly cancelled when we account for the threshold corrections in the coupling, as discussed in the next section. In detail, $\delta_{\chi}$ and $\delta_m$ remain unchanged, whereas we now have
\begin{eqnarray}
\delta_W &=& - \frac{\alpha_2}{4\pi} \left[ \frac{2n_D-3}{6\epsilon_{\rm UV}} + \frac{19-2n_D}{6\epsilon_{\rm IR}} + \frac{16}{3} L + \frac{16}{3} \ln 2 \right]\,, \nn
\delta_{g_2} &=& - \frac{\alpha_2}{4\pi} \left[ \frac{27-2n_D}{12\epsilon_{\rm UV}} \right]\,, \\
\delta_1 &=& - \frac{\alpha_2}{4\pi} \left[ \frac{4}{\epsilon_{\rm UV}} + \frac{67-2n_D}{12\epsilon_{\rm IR}} + \frac{44}{3} L + \frac{44}{3} \ln 2 + 8 \right]\,, \nonumber
\end{eqnarray}
so that in the Dirac fermion case
\begin{equation}\begin{aligned}
B_1^{[{\rm CT}]} &= \frac{\alpha_2}{4\pi} \left[ \frac{2n_D-43}{6\epsilon_{\rm IR}} - \frac{16}{3} L - \frac{16}{3} \ln 2  \right]\,, \\
B_2^{[{\rm CT}]} &= - \frac{1}{2} B_1^{[{\rm CT}]}\,, \\
B_3^{[{\rm CT}]} &= 0\,.
\end{aligned}\end{equation}

\subsection*{Threshold Corrections in the Coupling}

Throughout the above calculation we have treated the DM as a propagating degree of freedom and included its effects in loop diagrams. This implies that the coupling used so far above in this appendix implicitly depends on $n_D+1$ flavors -- $n_D$ left-handed SM doublets and one Majorana DM fermion -- i.e. we have used $\alpha_2 = \alpha_2^{(n_D+1)}(\mu)$. In the EFT however, the DM is no longer a propagating field and so the appropriate coupling is $\alpha_2^{(n_D)}(\mu)$. At order $\alpha_2^2$, which we are working to at one loop, the distinction will lead to a finite contribution because of the matching at the scale $\mu=m_{\chi}$, which we calculate in this section.

Let us start by reviewing the standard treatment of a running coupling in the $\overline{\rm MS}$ scheme. This running is captured by the $\beta$-function, which is defined by $\beta(\alpha_2) = \mu d\alpha_2/d\mu$, where here $\alpha_2$ is the renormalized coupling; the bare coupling is independent of $\mu$. The $\beta$-function can be written as:
\begin{equation}\begin{aligned}
\beta(\alpha_2) 
&= -2 \epsilon \alpha_2 - \frac{b_0}{2\pi} \alpha_2^2 + \ldots\,,
\end{aligned}\end{equation}
where we have expanded it to the order needed for this threshold matching analysis. At this order the LL solution for the running of the coupling is:
\begin{equation}
\alpha_2(\mu) = \frac{\alpha_2(\mu_0)}{1+\alpha_2(\mu_0) \frac{b_0}{2\pi} \ln \frac{\mu}{\mu_0} }\,.
\label{eq:coupling}
\end{equation}
In order to determine the threshold matching correction at the one-loop order we are working it suffices to simply demand that the coupling is continuous at the scale $m_{\chi}$, and this is captured by a difference in $b_0$. For our problem we define $b_0^{(n_D+1)}$ to be the value above $m_{\chi}$ and $b_0^{(n_D)}$ the value below. Then using Eq.~\eqref{eq:coupling} to define $\alpha_2^{(n_D+1)}(\mu)$ and $\alpha_2^{(n_D)}(\mu)$, it suffices to demand they match at a scale $m_{\chi}$, which gives:
\begin{equation}\begin{aligned}
&\alpha_2^{(n_D+1)}(\mu) = \alpha_2^{(n_D)}(\mu) \left[ 1 \vphantom{\frac{\mu}{m_{\chi}}} \right. \\
&\left. + \frac{\alpha_2^{(n_D)}(\mu)}{2\pi} \left( b_0^{(n_D)} - b_0^{(n_D+1)} \right) \ln \frac{\mu}{m_{\chi}} + \ldots \right]\,.
\label{eq:genalphamatch}
\end{aligned}\end{equation}
So now we just need to determine $b_0^{(n_D)} - b_0^{(n_D+1)}$. In general for a theory containing just gauge bosons, Weyl fermions (WF), Dirac fermions (DF), Majorana fermions (MF) and charged scalars (CS), we can write:
\begin{equation}\begin{aligned}
b_0 &= \frac{11}{3} C_A - \frac{2}{3} \sum_{i\in {\rm WF}} C(R_i) - \frac{4}{3} \sum_{i\in {\rm DF}} C(R_i) \\
&- \frac{2}{3} \sum_{i\in {\rm MF}} C(R_i) - \frac{1}{3} \sum_{i\in {\rm CS}} C(R_i)\,.
\end{aligned}\end{equation}
Our calculation can involve all of these ingredients: electroweak gauge bosons, the left-handed SM fermions (which are Weyl because only one chirality couples to the gauge bosons), the Majorana or Dirac DM fermion and the Higgs. Then using $C_A = 2$, $C(R) = 1/2$ for the SM left-handed fermions and the Higgs, and $C(R)=2$ for the adjoint Majorana Wino, we conclude:
\begin{equation}\begin{aligned}
b_0^{(n_D)} &= \frac{43-2n_D}{6}\,, \\
b_0^{(n_D+1)} &= \frac{35-2n_D}{6}\,.
\end{aligned}\end{equation}
From this Eq.~\eqref{eq:genalphamatch} tells us that to the order we are working:
\begin{equation}\begin{aligned}
\alpha_2^{(n_D+1)}(\mu) = \alpha_2^{(n_D)}(\mu) \left[ 1 + \frac{\alpha_2^{(n_D)}(\mu)}{4\pi} \left( \frac{8}{3} L + \frac{8}{3} \ln 2 \right) \right]\,.
\end{aligned}\end{equation}
Now as there is only a difference between the couplings at next to leading order, this only corrects the tree level result stated in Eq.~\eqref{eq:TreeLevelCoeff}. As such the impact of changing to the coupling defined below $m_{\chi}$, which is relevant for the matching, is to add the following contribution:
\begin{equation}\begin{aligned}
B_1^{[{\rm Matching}]} &= \frac{\alpha_2}{4\pi} \left[ \frac{8}{3} L + \frac{8}{3} \ln 2 \right]\,, \\
B_2^{[{\rm Matching}]} &= - \frac{1}{2} B_1^{[{\rm Matching}]}\,, \\
B_3^{[{\rm Matching}]} &= 0\,,
\label{eq:matchcontribution}
\end{aligned}\end{equation}
where after adding this contribution now here and in all earlier one-loop results we can simply take $\alpha_2 = \alpha_2^{(n_D)}$. 

In the Dirac case, the wino contributes differently to $b_0^{(n_D+1)}$, giving
\begin{equation}\begin{aligned}
b_0^{(n_D+1)} &= \frac{27-2n_D}{6}\,,
\end{aligned}\end{equation}
and accordingly we find
\begin{equation}\begin{aligned}
B_1^{[{\rm Matching}]} &= \frac{\alpha_2}{4\pi} \left[ \frac{16}{3} L + \frac{16}{3} \ln 2 \right]\,, \\
B_2^{[{\rm Matching}]} &= - \frac{1}{2} B_1^{[{\rm Matching}]}\,, \\
B_3^{[{\rm Matching}]} &= 0\,.
\label{eq:matchcontributionDirac}
\end{aligned}\end{equation}
Note that these corrections from threshold matching for the coupling exactly cancel the $L$ and $\ln 2$ terms that appear in $B_{1,2}^{\rm [CT]}$. 
As alluded to above, although the counterterm and threshold matching results individually differ for the Dirac or Majorana cases, when combined these terms , leading to the same result for the final $B_i^{(1)}$ coefficients given in Eq.~(\ref{eq:FullTheoryOneLoop}) below. 

\newpage
\begin{widetext}
\subsection*{Combination}
Combining the 25 graphs above with the counter-terms and the matching contributions, we arrive at the following result:
\begin{equation}\begin{aligned}
B_1^{(1)} &= \frac{\alpha_2}{4\pi} \left[ - \frac{4}{\epsilon^2} - \frac{48L+12i\pi-55-2n_D}{6\epsilon} - 8 L^2 - 4L - 4 i \pi L - 12 + \frac{11\pi^2}{6} \right]\,, \\
B_2^{(1)} &= \frac{\alpha_2}{4\pi} \left[ \frac{2}{\epsilon^2} + \frac{48L-12i\pi+79-2n_D}{12\epsilon} + 4L^2 + 6L - 2 i \pi L + 2 - \frac{5\pi^2}{12} \right]\,, \\
B_3^{(1)} &= \frac{\alpha_2}{4\pi} \left[ \frac{n_D - 72 \ln 2 - 71 + 3\pi^2}{12} \right]\,,
\label{eq:FullTheoryOneLoop}
\end{aligned}\end{equation}
where recall $L=\ln \mu/2m_{\chi}$, $n_D$ is the number of SM left-handed doublets and now all $\epsilon = \epsilon_{\rm IR}$. \\
\end{widetext}
As explained in detail at the outset of the calculation, the one-loop contribution to the matching coefficient is just the finite part of this result. Combining this with the tree-level term in Eq.~\eqref{eq:TreeLevelCoeff} and mapping back to $C_r$ using Eq.~\eqref{eq:CoeffMapping} then gives us the Wilson coefficients in Eq.~\eqref{eq:WilsonCoeff}, which we set out to justify.

If instead we had a Dirac DM triplet rather than a Majorana, then the only impact on the above would be for $B_3^{(1)}$, and we would instead have
\begin{equation}
B_3^{(1)} = \frac{\alpha_2}{4\pi} \left[ \frac{n_D - 72 \ln 2 - 43}{12} \right]\,.
\end{equation}

\section{Consistency Check on the High-Scale Matching}
\label{app:consistency}

As a non-trivial check on our high-scale calculation, we can calculate the $\ln \mu$, or $L$ in our case, pieces of Eq.~\eqref{eq:FullTheoryOneLoop} independently by expanding the NLL results. To begin with, if we define $C \equiv (C_1~C_2~C_3)^{T}$, then from the definition of the anomalous dimension we have:
\begin{equation}
\mu \frac{d}{d\mu} C(\mu) = \hat{\gamma}(\mu) C(\mu)\,.
\label{eq:DerivRunning}
\end{equation}
Next we expand the coefficients as a series in $\alpha_2$: $C(\mu) = C^{(0)}(\mu) + C^{(1)}(\mu) + ...$, where $C^{(0)}(\mu)$ is the tree-level contribution and $C^{(1)}(\mu)$ the one-loop result. Now we want a cross check on the one-loop contribution, so we evaluate Eq.~\eqref{eq:DerivRunning} at $\mathcal{O}(\alpha_2)$, giving
\begin{equation}
\mu \frac{d\alpha_2}{d\mu} \frac{\partial C^{(0)}}{\partial \alpha_2} + \mu \frac{\partial C^{(1)}(\mu)}{\partial \mu} = \hat{\gamma}_{\rm 1-loop}(\mu) C^{(0)}(\mu)\,,
\label{eq:DerivRunningorder2}
\end{equation}
and rearranging we arrive at:
\begin{equation}
\mu \frac{\partial C^{(1)}(\mu)}{\partial \mu} = \hat{\gamma}_{\rm 1-loop}(\mu) C^{(0)}(\mu) - \mu \frac{d\alpha_2}{d\mu} \frac{\partial C^{(0)}}{\partial \alpha_2}\,.
\label{eq:Consistency}
\end{equation}
This equation shows that we can derive the $\mu$ and hence $L$ dependence of the one-loop Wilson coefficient from the one-loop anomalous dimension and tree-level Wilson coefficient, both of which are known from the NLL result. To be more explicit, we can write the bare Wilson coefficient as
\begin{equation}\begin{aligned}
C_{\rm bare} = &\mu^{2\epsilon} \left( \frac{a}{\epsilon^2} + \frac{b}{\epsilon} + \mu {\rm -independent} \right) \\
= &\frac{a}{\epsilon^2} + \frac{b+2a L}{\epsilon} + 2 a L^2 + 2 b L \\
&+ \mu {\rm -independent}\,,
\end{aligned}\end{equation}
where in the second equality we swapped frpm $\ln \mu$ to $L$ and absorbed the additional $\ln 2$ factors into the $\mu$-independent term. From here we can write the renormalized Wilson coefficient as
\begin{equation}
C_{\rm ren.} = 2 a L^2 + 2 b L + \mu {\rm -independent}\,,
\end{equation}
which we can then substitute into the left-hand side of Eq.~\eqref{eq:Consistency} to derive $a$ and $b$ for each Wilson coefficient. Doing this and then mapping back to $B_r$ using Eq.~\eqref{eq:CoeffMapping}, we find
\begin{eqnarray}
B_1^{(1)} &=& \frac{\alpha_2}{4\pi} \left[ - \frac{8L}{\epsilon} - 8L^2 -4L - 4i \pi L + \mu {\rm -ind.} \right]\,, \nn
B_2^{(1)} &=& \frac{\alpha_2}{4\pi} \left[ \frac{4L}{\epsilon} + 4L^2 + 6L - 2i \pi L + \mu {\rm -ind.} \right]\,, \nn
B_3^{(1)} &=& \frac{\alpha_2}{4\pi} \left[ 0 + \mu {\rm -ind.} \right]\,,
\end{eqnarray}
in exact agreement with Eq.~\eqref{eq:FullTheoryOneLoop}. In particular, as $B_3^{(0)}=0$, we needed $B_3^{(1)}$ to be independent of $L$, as we found.

\newpage
\section{Low-Scale Matching Calculation}
\label{app:lowscalematching}

The focus of this appendix is to derive the low-scale matching conditions stated in Eqs.~\eqref{eq:lowbreakdown}, \eqref{eq:LowSoft}, \eqref{eq:LowColinear}, \eqref{eq:LowColinearConsts1}, and \eqref{eq:LowColinearConsts2}. At this scale, the matching is from an effective theory where the $W$, $Z$, top and Higgs are dynamical degrees of freedom -- NRDM-SCET$_{\rm EW}$ -- onto a theory where these electroweak modes have been integrated out -- NRDM-SCET$_{\gamma}$.

In order to perform the calculation we will make use of the formalism of electroweak SCET developed in \cite{Chiu:2007yn,Chiu:2007dg,Chiu:2008vv,Chiu:2009mg,Chiu:2009ft}. As we are working in SCET, there are both collinear and soft gauge boson diagrams that will appear in the one-loop matching. In \cite{Chiu:2009mg} it was proven that at one-loop the total low-scale matching contribution from these soft and collinear SCET modes can always be decomposed into a contribution that is diagonal, in that it leads to no operator mixing, and another that is non-diagonal, as it does induce mixing. In their works, they then refer to the diagonal parts as \textit{collinear} and non-diagonal ones as \textit{soft}, however we shall always use the term ``diagonal" to refer to the contributions that have contributions from both soft and collinear diagrams, although we do use a subscript ``c" for the diagonal piece. At one loop the matching amounts to evaluating the diagrams that appear in NRDM-SCET$_{\rm EW}$ but not NRDM-SCET$_{\gamma}$. These diagrams can be broken into three classes:
\begin{enumerate}
\item Wave-function diagrams correcting our initial non-relativistic states;
\item Diagrams where a soft gauge boson is exchanged between two different external states; and
\item Final state collinear diagrams, which are now corrections to collinear states.
\end{enumerate}
Each class will be discussed separately below. Before doing so, however, we first define our operators and outline how the low-scale matching proceeds at tree level.

Unlike for the high-scale matching, here we only consider the two operators that match onto $\mathcal{M}_A$ in Eq.~\eqref{eq:FullTheoryOps}, as opposed to the third operator coming from $\mathcal{M}_B$. The reason for this is the additional operator does not contribute to the low-scale matching calculation for present day DM annihilation at any order in leading power NRDM-SCET. To understand this note that the operators coming from $\mathcal{M}_A$ and $\mathcal{M}_B$ have different spin structures. In order to mix these structures we need to transfer angular momentum between the states. The only low-scale graphs we can write down to do this are soft gauge boson exchanges. The spin structure of the coupling of a soft exchange to an $n$-collinear gauge boson is $\slashed{n}$ and the corresponding coupling to our non-relativistic DM field is $\slashed{v}$. Neither coupling allows for a transfer of angular momentum, demonstrating that these operators cannot mix. Unlike for the high-scale matching, we will not make use of the operator corresponding to $\mathcal{M}_B$ for our low-scale consistency check, so we drop it from consideration at the outset.

\subsection*{Operator Definition and Tree-level Matching}

Prior to electroweak symmetry breaking, the two relevant operators in NRDM-SCET$_{\rm EW}$ can be written schematically as:
\begin{equation}\begin{aligned}
\mathcal{O}_1 &= \frac{1}{2} \delta_{ab} \delta_{cd} \chi^a \chi^b W_3^c W_4^d\,, \\
\mathcal{O}_2 &= \frac{1}{4} \left( \delta_{ac} \delta_{bd} + \delta_{ad} \delta_{bc} \right) \chi^a \chi^b W_3^c W_4^d\,.
\label{eq:UnbrokenOps}
\end{aligned}\end{equation}
Our notation here is schematic in the sense that we have suppressed the Lorentz structure and soft Wilson lines. The form of these is written out explicitly in Eq.~\eqref{eq:Ops} and is left out for convenience as it appears in every operator written down in this appendix. Further, in this equation the factor of $1/2$ is introduced for convenience; as $\chi$ is a Majorana field this factor ensures the Feynman rule associated with these operators has no additional numerical factor. Note also that the gauge bosons are labelled as they are associated with a collinear direction. At tree-level the low-scale matching is effected simply by mapping the fields in these operators onto their broken form. Explicitly we have:
\begin{equation}\begin{aligned}
\chi^1 &= \frac{1}{\sqrt{2}} \left( \chi^+ + \chi^- \right)\,, \\
\chi^2 &= \frac{i}{\sqrt{2}} \left( \chi^+ - \chi^- \right)\,, \\
\chi^3 &= \chi^0\,, \\
W^1 &= \frac{1}{\sqrt{2}} \left( W^+ + W^- \right)\,, \\
W^2 &= \frac{i}{\sqrt{2}} \left( W^+ - W^- \right)\,, \\
W^3 &= s_W A + c_W Z\,.
\label{eq:BrokenFields}
\end{aligned}\end{equation}
Substituting these into Eq.~\eqref{eq:UnbrokenOps} yields 22 operators in the broken theory. Of these, 14 involve a $W^{\pm}$ in the final state, so we will not consider them further. We define the remaining 8 as:
\begin{equation}\begin{aligned}
\hat{\mathcal{O}}_1 &= \frac{1}{2} \chi^0 \chi^0 A_3 A_4\,,\hspace{0.2in}\hat{\mathcal{O}}_2 = \frac{1}{2} \chi^0 \chi^0 Z_3 A_4\,,\\
\hat{\mathcal{O}}_3 &= \frac{1}{2} \chi^0 \chi^0 A_3 Z_4\,,\hspace{0.21in}\hat{\mathcal{O}}_4 = \frac{1}{2} \chi^0 \chi^0 Z_3 Z_4\,,\\
\hat{\mathcal{O}}_5 &= \chi^+ \chi^- A_3 A_4\,,\hspace{0.245in}\hat{\mathcal{O}}_6 = \chi^+ \chi^- Z_3 A_4\,,\\
\hat{\mathcal{O}}_7 &= \chi^+ \chi^- A_3 Z_4\,,\hspace{0.255in}\hat{\mathcal{O}}_8 = \chi^+ \chi^- Z_3 Z_4\,,
\label{eq:BrokenOps}
\end{aligned}\end{equation}
where again we have used the schematic notation of Eq.~\eqref{eq:UnbrokenOps}, as we will for all operators in this appendix. At tree level, the operators in Eq.~\eqref{eq:UnbrokenOps} and \eqref{eq:BrokenOps} are related simply by the change of variables in Eq.~\eqref{eq:BrokenFields}. This mapping is performed by a $22 \times 2$ matrix, but again we only state the part of this matrix we are interested in:
\begin{equation}
\hat{D}_{s,1-8}^{(0)} = \begin{bmatrix} 
s_W^2 & s_W^2 \\
s_W c_W & s_W c_W \\
s_W c_W & s_W c_W \\
c_W^2 & c_W^2 \\
s_W^2 & 0 \\
s_W c_W & 0 \\
s_W c_W & 0 \\
c_W^2 & 0 
\end{bmatrix}\,.
\label{eq:TreeLevelMapping}
\end{equation}
In terms of the calculation presented in the main text, what we actually want is the mapping onto the Sudakov factors $\Sigma$, defined in Eq.~\eqref{eq:Factorized}, not the broken operators in Eq.~\eqref{eq:BrokenOps}. As given there, the $s_W$ and $c_W$ factors are absorbed into $P_X$, and so will not contribute to the $\Sigma$ factors. Then $\hat{\mathcal{O}}_{1-4}$ represent the contributions to neutral annihilation $\chi^0 \chi^0 \to X$, represented by $\Sigma_1 - \Sigma_2$, and $\hat{\mathcal{O}}_{5-8}$ the contributions to charged annihilation $\chi^+ \chi^- \to X$, represented by $\Sigma_1$. Accordingly we have:
\begin{equation}
\hat{D}^{(0)}_s = \begin{bmatrix}
1 & 0 \\
1 & 1
\end{bmatrix}\,.
\label{eq:TreeLevelLowScale}
\end{equation}
This provides the tree-level result we should use in Eq.~\eqref{eq:lowbreakdown}. Next we turn to calculating this one-loop low-scale matching in full, considering the three classes of diagrams that can contribute in turn.

\subsection*{Initial State Wave-function Graphs}

There are two graphs that fall under the category of initial state wave-function corrections, and these are shown below.
\begin{center}
\includegraphics[height=0.2\columnwidth]{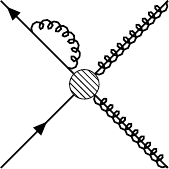} \hspace{0.1in}
\includegraphics[height=0.2\columnwidth]{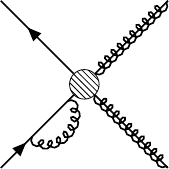}
\end{center}
Note here we follow the standard SCET conventions of drawing collinear fields as gluons with a solid line through them, whereas soft fields are represented simply by gluon lines. In these graphs, the soft gauge field can be either a $W$ or $Z$ boson. In either case the integral to be calculated is:
\begin{equation}
- g^2 \int \dbar^d k \frac{\mu^{2\epsilon}}{[k^2-m^2]v \cdot (k + p)}\,,
\end{equation}
where $g$ is the coupling -- $g_2$ for a $W$ boson, $c_W g_2$ for a $Z$ boson, $p$ is the external momentum, $k$ is the loop momentum, $m$ the gauge boson mass, and $v$ is the velocity associated with the non-relativistic $\chi$ field. Given our initial state is heavy, this is unsurprisingly exactly the heavy quark effective theory wave-function renormalization graph. The analytic solution can be found in e.g. \cite{Stewart:1998ke,Manohar:2000dt}, and using this we find:
\begin{equation}
= - i v \cdot p \frac{\alpha}{2\pi} \left[ \frac{1}{\epsilon} + \ln \frac{\mu^2}{m^2} \right]\,,
\end{equation}
where $\alpha = g^2/4\pi$. Now in addition to the one-loop graphs we drew above, at this order there will also be a counter-term of the form $i v \cdot p (Z_{\chi} - 1)$. Again working in the on-shell scheme so that we do not need to consider the residues, we conclude:
\begin{equation}
Z_{\chi} = 1 + \frac{\alpha_2(\mu)}{2\pi} \left[ \frac{1}{\epsilon} - \ln \frac{m_W^2}{\mu^2} - c_W^2 \ln \frac{m_Z^2}{\mu^2} \right]\,.
\end{equation}
Now each of our initial states will contribute $Z_{\chi}^{1/2}$, implying that the contribution to $\hat{D}(\mu)$ given in Eq.~\eqref{eq:lowbreakdown} is
\begin{equation}
D_c^{\chi}(\mu) = 1 - \frac{\alpha_2(\mu)}{2\pi} \left[\ln \frac{m_W^2}{\mu^2} + c_W^2 \ln \frac{m_Z^2}{\mu^2} \right]\,,
\label{eq:iswfgderivation}
\end{equation}
and the subscript $c$ indicates this is a diagonal contribution in the sense that it leads to no operator mixing. This is exactly as in Eq.~\eqref{eq:LowColinear} and justifies this part of the low-scale matching.

\subsection*{Soft Gauge Boson Exchange Graphs}

In this section we calculate the contribution from the exchange of a soft $W$ or $Z$ gauge boson between different external final states. As these gauge bosons carry SU(2)$_{\rm L}$ gauge indices, unsurprisingly these graphs will lead to operator mixing. Consequently, in terms of the notation introduced above these graphs will lead to non-diagonal contributions. They will also induce diagonal terms, and we will carefully separate the two below.

Once separated, we will group the diagonal contribution with those we get from the final state wave-function graphs we consider in the next subsection. The reason for this is that these diagonal contributions for photon and $Z$ final states, as we have, were already evaluated in \cite{Chiu:2009ft}, and we will not fully recompute them here. In that work, however, the diagonal contribution was only stated in full. The breakdown into the soft boson exchange and final state wave-function graphs was not provided. This raises a potential issue because in that work all external states were taken to be collinear, not non-relativistic. As such, in this section we will explicitly calculate the soft gauge boson exchange graphs for both kinematics and demonstrate that the diagonal contribution is identical in the two cases.

Before calculating the graphs, we first introduce some useful notation. At one loop the gauge bosons will have two couplings to the four external states. Each of these couplings will have an associated gauge index structure, and in order to deal with this it is convenient to introduce gauge index or color operators $\mathbf{T}$. This notation was first introduced in \cite{Catani:1996jh,Catani:1996vz}, and it allows the gauge index structure to be organized generally rather than case by case. Examples can be found in the original papers and also in the SCET literature e.g. \cite{Chiu:2009mg,Chiu:2009ft,Moult:2015aoa}. An example relevant for our purposes is the action of $\mathbf{T}$ on an SU(2)$_{\rm L}$ adjoint, which is the representation of both our initial and final states:
\begin{equation}\begin{aligned}
\mathbf{T} \chi^a &= (T_A^c)_{a a^{\prime}} \chi^{a^{\prime}} = - i\epsilon_{ca a^{\prime}} \chi^{a^{\prime}}\,, \\
\mathbf{T} W^a &= (T_A^c)_{a a^{\prime}} W^{a^{\prime}} = - i\epsilon_{ca a^{\prime}} W^{a^{\prime}}\,.
\label{eq:gaugeindexopaction}
\end{aligned}\end{equation}
In terms of this notation then, we can write the gauge index structure of all relevant one-loop low-scale matching graphs as $\mathbf{T}_i \cdot \mathbf{T}_j$, where $i,j$ label any of the four external legs. Because of this we label the result from these soft exchange diagrams as $S_{ij}$ for the case of our kinematics -- non-relativistic initial states and collinear final states -- and we use $S^{\prime}_{ij}$ to denote the kinematics of \cite{Chiu:2009ft} -- all external states collinear. Following \cite{Chiu:2009mg,Chiu:2009ft}, we take all external momenta to be incoming and further rapidity divergences will be regulated with the $\Delta$-regulator \cite{Chiu:2009yx}. Now let us turn to the graphs one by one.

\subsection*{\large $S^{(\prime)}_{12}$}

\begin{center}
\includegraphics[height=0.2\columnwidth]{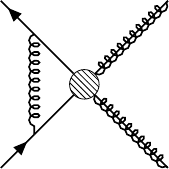}
\end{center}
In this graph the soft gauge boson exchanged between the initial state can be a $W$ or $Z$ boson. In either case, the value of this graph is:
\begin{eqnarray}
S_{12} &=& \frac{\alpha}{2\pi} \mathbf{T}_1 \cdot \mathbf{T}_2 \left[ \frac{1}{\epsilon} - \ln \frac{m^2}{\mu^2} \right]\,, \label{eq:S12} \\
S^{\prime}_{12} &=& \frac{\alpha}{2\pi} \mathbf{T}_1 \cdot \mathbf{T}_2 \left[ \frac{1}{\epsilon^2} - \frac{1}{\epsilon} \left( \ln \frac{\delta_1 \delta_2}{\mu^2} + i \pi \right) - \frac{1}{2} \ln^2 \frac{m^2}{\mu^2}  \right. \nn
&&\hspace{0.83in} \left.+ i \pi \ln \frac{m^2}{\mu^2} + \ln \frac{m^2}{\mu^2} \ln \frac{\delta_1 \delta_2}{\mu^2} - \frac{\pi^2}{12} \right]\,, \nonumber
\end{eqnarray}
where as above $\alpha = g^2/4\pi$ and the identity $g$ and $m$ depend on whether this is for a $W$ or $Z$. In $S^{\prime}_{12}$, $\delta_{1}$ and $\delta_2$ are the $\Delta$-regulators. Unsurprisingly these only appear for the collinear kinematics for the initial state in $S^{\prime}_{12}$ and not for the nonrelativistic kinematics in $S_{12}$ .

\subsection*{\large $S^{(\prime)}_{13}$, $S^{(\prime)}_{14}$, $S^{(\prime)}_{23}$, and $S^{(\prime)}_{24}$}

\begin{center}
\includegraphics[height=0.2\columnwidth]{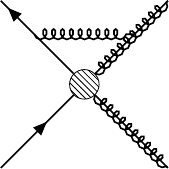} \hspace{0.1in}
\includegraphics[height=0.2\columnwidth]{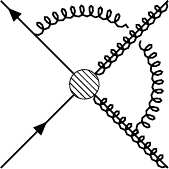} \\
\vspace{0.1in}
\includegraphics[height=0.2\columnwidth]{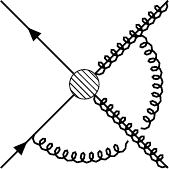} \hspace{0.1in}
\includegraphics[height=0.2\columnwidth]{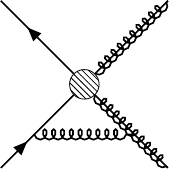}
\end{center}
Again the exchanged soft boson can be a $W$ or $Z$. These four graphs are grouped together as they have a common form, for example:
\begin{eqnarray}
S_{13} &=& \frac{\alpha}{2\pi} \mathbf{T}_1 \cdot \mathbf{T}_3 \left[ \frac{1}{2\epsilon^2} - \frac{1}{2\epsilon} \ln \frac{\delta_3^2}{\mu^2} - \frac{1}{4} \ln^2 \frac{m^2}{\mu^2} \right. \label{eq:S13} \\
&&\hspace{1.00in}\left.+ \frac{1}{2} \ln \frac{\delta_3^2}{\mu^2} \ln \frac{m^2}{\mu^2} - \frac{\pi^2}{24} \right]\,, \nn
S_{13}^{\prime} &=& \frac{\alpha}{2\pi} \mathbf{T}_1 \cdot \mathbf{T}_3 \left[ \frac{1}{\epsilon^2} - \frac{1}{\epsilon} \ln \left( - \frac{\delta_1 \delta_3}{\mu^2 w_{13}} \right) - \frac{1}{2} \ln^2 \frac{m^2}{\mu^2} \right. \nn
&&\hspace{0.97in}\left.+ \ln \frac{m^2}{\mu^2} \ln \left( - \frac{\delta_1 \delta_3}{\mu^2 w_{13}} \right) - \frac{\pi^2}{12} \right]\,. \nonumber
\end{eqnarray}
Then $S^{(\prime)}_{14}$ is given by the same expressions but with $3 \to 4$, whilst $S^{(\prime)}_{23}$ and $S^{(\prime)}_{24}$ are given by similar replacements. For the all collinear case we have defined the following functions of the kinematics:
\begin{equation}\begin{aligned}
w_{13} &= w_{24} \equiv \frac{1}{2} n_1 \cdot n_3 = \frac{1}{2} n_2 \cdot n_4 = \frac{t}{s}\,, \\
w_{14} &= w_{23} \equiv \frac{1}{2} n_1 \cdot n_4 = \frac{1}{2} n_2 \cdot n_3 = \frac{u}{s}\,,
\label{eq:wkin}
\end{aligned}\end{equation}
where $s$, $t$, and $u$ are the Mandelstam variables relevant for all incoming momenta. The signs inside the logs in Eq.~\eqref{eq:S13} can be understood by noting that as $t < 0$, $u < 0$, whilst $s > 0$, we have $w_{ij} < 0$.

\subsection*{\large $S^{(\prime)}_{34}$}

\begin{center}
\includegraphics[height=0.2\columnwidth]{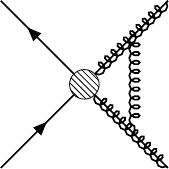}
\end{center}
Finally we have the graph above, which yields:
\begin{eqnarray}
S_{34} &=& \frac{\alpha}{2\pi} \mathbf{T}_3 \cdot \mathbf{T}_4 \left[ \frac{1}{\epsilon^2} - \frac{1}{\epsilon} \left( \ln \frac{\delta_3 \delta_4}{\mu^2} + i \pi \right) - \frac{1}{2} \ln^2 \frac{m^2}{\mu^2} \right. \nn
&&\hspace{0.83in} \left.+ i \pi \ln \frac{m^2}{\mu^2} + \ln \frac{m^2}{\mu^2} \ln \frac{\delta_3 \delta_4}{\mu^2} - \frac{\pi^2}{12} \right]\,, \nn
S_{34}^{\prime} &=& S_{34}\,.
\label{eq:S34}
\end{eqnarray}

This completes the list of graphs to evaluate. As written it appears that all graphs are non-diagonal from their gauge index structure. However as we will now show, the combinations of all graphs can be reduced to a diagonal and non-diagonal piece. Firstly for the case of all collinear external states we have:
\begin{equation}
S^{\prime}_{12} + S^{\prime}_{13} + S^{\prime}_{14} + S^{\prime}_{23} + S^{\prime}_{24} + S^{\prime}_{34} \equiv \sum_{\langle i j \rangle} S^{\prime}_{ij}\,,
\label{eq:angsum}
\end{equation}
which serves to define $\langle i j \rangle$. The part of this sum that involving the rapidity regulators can be written as
\begin{equation}
\frac{\alpha}{2\pi} \ln \frac{m^2}{\mu^2} \sum_{\langle i j \rangle} \mathbf{T}_i \cdot \mathbf{T}_j \left( \ln \frac{\delta_i}{\mu} + \ln \frac{\delta_j}{\mu} \right)\,.
\label{eq:acdeltareg}
\end{equation}
This can be simplified using the following identity:\footnote{This and the gauge index identity stated below in Eq.~\eqref{eq:gaugeindexid2} follow simply from the fact $\sum_i \mathbf{T}_i = 0$ when it acts on gauge index singlet operators, see for example \cite{Chiu:2009mg}.}
\begin{equation}
\sum_{\langle i j \rangle } \left( f_i + f_j \right) \mathbf{T}_i \cdot \mathbf{T}_j = - \sum_i f_i \mathbf{T}_i \cdot \mathbf{T}_i\,.
\label{eq:gaugeindexid1}
\end{equation}
If we identify $f_i = \ln \delta_i / \mu$, then Eq.~\eqref{eq:acdeltareg} becomes:
\begin{equation}
= - \frac{\alpha}{2\pi} \ln \frac{m^2}{\mu^2} \sum_{\langle i j \rangle} \mathbf{T}_i \cdot \mathbf{T}_i \ln \frac{\delta_i}{\mu} \,,
\label{eq:acdeltaregsimp}
\end{equation}
which is now diagonal in the gauge indices. For the remaining terms that are independent of $\delta$, we organise them as follows:
\begin{equation}\begin{aligned}
\sum_{\langle i j \rangle} S^{\prime}_{ij}
&= \frac{1}{2} \left[ S^{\prime}_{12} + S^{\prime}_{13} + S^{\prime}_{14} \right] \\
&+ \frac{1}{2} \left[ S^{\prime}_{21} + S^{\prime}_{23} + S^{\prime}_{24} \right] \\
&+ \frac{1}{2} \left[ S^{\prime}_{31} + S^{\prime}_{32} + S^{\prime}_{34} \right] \\
&+ \frac{1}{2} \left[ S^{\prime}_{41} + S^{\prime}_{42} + S^{\prime}_{43} \right] \,,
\label{eq:gaugeindexgboost}
\end{aligned}\end{equation}
where we used the fact $S^{\prime}_{ij} = S^{\prime}_{ji}$. Each of these groups can now be simplified. For example, the first group can be written as:
\begin{eqnarray}
S^{\prime}_{12} + S^{\prime}_{13} + S^{\prime}_{14} &= &\frac{\alpha}{2\pi} \left( \mathbf{T}_1 \cdot \mathbf{T}_2 + \mathbf{T}_1 \cdot \mathbf{T}_3 + \mathbf{T}_1 \cdot \mathbf{T}_4 \right) \nn
&&\times \left[ - \frac{1}{2} \ln^2 \frac{m^2}{\mu^2} - \frac{\pi^2}{12} \right] \nn
&&+ \frac{\alpha}{2\pi} \mathbf{T}_1 \cdot \mathbf{T}_2 \left[ i \pi \ln \frac{m^2}{\mu^2} \right] \label{eq:boostorg} \\
&&- \frac{\alpha}{2\pi} \mathbf{T}_1 \cdot \mathbf{T}_3 \left[ \ln \left( - \frac{t}{s} \right) \ln \frac{m^2}{\mu^2} \right] \nn
&&- \frac{\alpha}{2\pi} \mathbf{T}_1 \cdot \mathbf{T}_4 \left[ \ln \left( - \frac{u}{s} \right) \ln \frac{m^2}{\mu^2} \right]\,, \nonumber
\end{eqnarray}
If we then use
\begin{equation}
\sum_{j,j\neq i} \mathbf{T}_i \cdot \mathbf{T}_j = - \mathbf{T}_i \cdot \mathbf{T}_i\,,
\label{eq:gaugeindexid2}
\end{equation}
Eq.~\eqref{eq:boostorg} can be rewritten as:
\begin{eqnarray}
&=&\frac{\alpha}{2\pi} \mathbf{T}_1 \cdot \mathbf{T}_1 \left[ \frac{1}{2} \ln^2 \frac{m^2}{\mu^2} + \frac{\pi^2}{12} \right] \nn
&&+ \frac{\alpha}{2\pi} \mathbf{T}_1 \cdot \mathbf{T}_2 \left[ i \pi \ln \frac{m^2}{\mu^2} \right] \label{eq:boostorg2} \\
&&- \frac{\alpha}{2\pi} \mathbf{T}_1 \cdot \mathbf{T}_3 \left[ \ln \left( - \frac{t}{s} \right) \ln \frac{m^2}{\mu^2} \right] \nn
&&- \frac{\alpha}{2\pi} \mathbf{T}_1 \cdot \mathbf{T}_4 \left[ \ln \left( - \frac{u}{s} \right) \ln \frac{m^2}{\mu^2} \right]\,. \nonumber
\end{eqnarray}
Repeating this for the remaining three terms in Eq.~\eqref{eq:gaugeindexgboost} and reinserting the $\delta$ contributions, we can rewrite the combination of all terms as:
\begin{equation}
\sum_{\langle i j \rangle} S^{\prime}_{ij} \equiv \sum_{\langle i j \rangle} \hat{S}^{\prime}_{ij} + \sum_i C_i\,,
\label{eq:boostreduce}
\end{equation}
where we have defined:
\begin{eqnarray}
\hat{S}^{\prime}_{ij} &\equiv& - \frac{\alpha}{2\pi} \ln \frac{m^2}{\mu^2} \mathbf{T}_i \cdot \mathbf{T}_j U_{ij}^{\prime}\,, \label{eq:boostsoftcollinear} \\
C_i &\equiv& \frac{\alpha}{2\pi} \mathbf{T}_i \cdot \mathbf{T}_i \left[ \frac{1}{4} \ln^2 \frac{m^2}{\mu^2} + \frac{\pi^2}{24} - \frac{1}{2} \ln \frac{m^2}{\mu^2} \ln \frac{\delta_i^2}{\mu^2} \right]\,, \nonumber
\end{eqnarray}
and from the above we can see that:
\begin{equation}\begin{aligned}
U_{12}^{\prime} &= U_{34}^{\prime} = - i \pi\,, \\
U_{13}^{\prime} &= U_{24}^{\prime} = \ln \left( - \frac{t}{s} \right)\,, \\
U_{14}^{\prime} &= U_{23}^{\prime} = \ln \left( - \frac{u}{s} \right)\,.
\label{eq:boostUij}
\end{aligned}\end{equation}
Thus as claimed, we have reduced $\sum_{\langle i j \rangle} S^{\prime}_{ij}$ in Eq.~\eqref{eq:boostreduce} into a diagonal and non-diagonal piece. Importantly we have explicitly isolated the diagonal contribution $C_i$, and as we will now show we get exactly the same diagonal contribution for the kinematics of interest in this work.

Before doing so, however, note that the irreducibly non-diagonal contribution given in Eq.~\eqref{eq:boostsoftcollinear} and Eq.~\eqref{eq:boostUij} agrees with Eq.~(150) in \cite{Chiu:2009mg}, where they gave the general form of $U_{ij}^{\prime}$ for the case of all external collinear particles:
\begin{equation}
U_{ij}^{\prime} = \ln \frac{-n_i \cdot n_j - i0^+}{2}\,.
\end{equation}

Next we repeat this procedure for $\sum_{\langle i j \rangle} S_{ij}$, where we have non-relativistic fields in the initial state. As before we consider the contribution from the rapidity regulators at the outset, which for $\delta_3$ yield:
\begin{eqnarray}
&&\frac{\alpha}{2\pi} \left( \mathbf{T}_1 \cdot \mathbf{T}_2 + \mathbf{T}_1 \cdot \mathbf{T}_3 + \mathbf{T}_1 \cdot \mathbf{T}_4 \right) \left[ \frac{1}{2} \ln \frac{m^2}{\mu^2} \ln \frac{\delta_3^2}{\mu^2} \right] \nn
&=&- \frac{\alpha}{2\pi} \mathbf{T}_3 \cdot \mathbf{T}_3 \left[ \frac{1}{2} \ln \frac{m^2}{\mu^2} \ln \frac{\delta_3^2}{\mu^2} \right]\,,
\label{eq:nrdeltaorg}
\end{eqnarray}
where we again used Eq.~\eqref{eq:gaugeindexid2}. An identical relation will hold for $\delta_4$, and this time there is no $\delta_{1}$ or $\delta_2$ as the non-relativistic fields do not lead to rapidity divergences. For the remaining terms, we now organise them as follows:
\begin{equation}\begin{aligned}
\sum_{\langle i j \rangle} S_{ij}
= S_{12} 
+ &\left[ S_{31} + S_{32} + \frac{1}{2} S_{34} \right] \\
+ &\left[ S_{41} + S_{42} + \frac{1}{2} S_{43} \right]\,.
\label{eq:gaugeindexgnr}
\end{aligned}\end{equation}
Evaluating each of the terms in square brackets and simplifying the gauge index structure as before, we arrive at the following:
\begin{equation}
\sum_{\langle i j \rangle} S_{ij} \equiv \sum_{\langle i j \rangle} \hat{S}_{ij} + C_3 + C_4\,,
\label{eq:nrreduce}
\end{equation}
where we again have:
\begin{eqnarray}
\hat{S}_{ij} &\equiv& - \frac{\alpha}{2\pi} \ln \frac{m^2}{\mu^2} \mathbf{T}_i \cdot \mathbf{T}_j U_{ij}\,, \label{eq:nrsoftcollinear} \\
C_i &\equiv& \frac{\alpha}{2\pi} \mathbf{T}_i \cdot \mathbf{T}_i \left[ \frac{1}{4} \ln^2 \frac{m^2}{\mu^2} + \frac{\pi^2}{24} - \frac{1}{2} \ln \frac{m^2}{\mu^2} \ln \frac{\delta_i^2}{\mu^2} \right]\,, \nonumber
\end{eqnarray}
and now
\begin{equation}\begin{aligned}
U_{12} &= 1\,, \\
U_{34} &= - i \pi\,, \\
U_{13} &= U_{24} = U_{14} = U_{23} = 0 \,.
\label{eq:noboostUij}
\end{aligned}\end{equation}
Critically, although the non-diagonal contribution is different to the case of all collinear kinematics, we see that the diagonal function defined in Eq.~\eqref{eq:nrsoftcollinear} is identical to that in Eq.~\eqref{eq:boostsoftcollinear}. This justifies the claim made earlier that the diagonal part of this calculation is the same for both kinematics. As such we put the $C_i$ terms aside for the moment, and return to them when we consider the final state collinear graphs.

What remains here then is to evaluate the irreducibly non-diagonal contribution: $\sum_{\langle i j \rangle} \hat{S}_{ij}$. This essentially amounts to calculating the gauge index structure, which the use of gauge index operators has allowed us to put off until now. In addition we need to recall that we have a contribution to each graph from a $W$ and $Z$ boson exchange. As above we closely follow the approach in \cite{Chiu:2009mg,Chiu:2009ft}, except accounting for the differences in our kinematics. To this end, we begin by observing that after electroweak symmetry breaking the unbroken SU(2)$_{\rm L}$ and U(1)$_Y$ generators, $\mathbf{t}$ and $Y$, become
\begin{equation}\begin{aligned}
\alpha_2 \mathbf{t} \cdot \mathbf{t} + \alpha_1 Y \cdot Y \to &\frac{1}{2} \alpha_W (t_+ t_- + t_- t_+) \\
&+ \alpha_Z t_Z \cdot t_Z + \alpha_{\rm em} Q \cdot Q\,,
\end{aligned}\end{equation}
where $\alpha_2 = \alpha_{\rm em}/s_W^2$, $\alpha_1 = \alpha_{\rm em}/c_W^2$, $\alpha_W = \alpha_2$, $\alpha_Z = \alpha_2/c_W^2$, and $t_Z = t_3 - s_W^2 Q$. This implies that we can write the full contribution as:
\begin{eqnarray}
\hat{D}_s^{(1)} &= &\frac{\alpha_W(\mu)}{2\pi} \ln \frac{m_W^2}{\mu^2} \left[ - \sum_{\langle i j \rangle} \frac{1}{2} (t_+ t_- + t_- t_+) U_{ij} \right] \nn
&&+ \frac{\alpha_Z(\mu)}{2\pi} \ln \frac{m_Z^2}{\mu^2} \left[ - \sum_{\langle i j \rangle} t_{Zi} t_{Zj} U_{ij} \right]\,.
\end{eqnarray}
Now the contribution on the first line is more complicated, because $(t_+ t_- + t_- t_+)U_{ij}$ is a non-diagonal $22 \times 22$ matrix, whereas as we will see $t_{Zi} t_{Zj} U_{ij}$ is diagonal. Nevertheless we can simplify the non-diagonal part by using the following relation:
\begin{equation}\begin{aligned}
\frac{1}{2} (t_+ t_- + t_- t_+) = \mathbf{t} \cdot \mathbf{t} - t_3 \cdot t_3\,.
\end{aligned}\end{equation}
Here $t_3 \cdot t_3$ is again diagonal, and whilst $\mathbf{t} \cdot \mathbf{t}$ is non-diagonal, it is written in terms of the unbroken operators so that we can calculate it in the unbroken theory where we only have 2 operators not 22. Thus it is now a $2 \times 2$ matrix. In terms of this we can now write the non-diagonal contribution to the low-scale matching as:
\begin{equation}\begin{aligned}
\hat{D}_s &= \hat{D}_s^{(0)} + \hat{D}_{s,W}^{(1)} + \hat{D}_{s,Z}^{(1)}\,, \\
\hat{D}_{s,W}^{(1)} &= \frac{\alpha_W(\mu)}{2\pi} \ln \frac{m_W^2}{\mu^2} \left[ \hat{D}_s^{(0)} \mathfrak{S} + \mathfrak{D}_W \hat{D}_s^{(0)} \right]\,, \\
\hat{D}_{s,Z}^{(1)} &= \frac{\alpha_Z(\mu)}{2\pi} \ln \frac{m_Z^2}{\mu^2} \left[ \mathfrak{D}_Z \hat{D}_s^{(0)} \right]\,,
\label{eq:ndlsfulldef}
\end{aligned}\end{equation}
where $\hat{D}_s^{(0)}$ is given in Eq.~\eqref{eq:TreeLevelMapping} and as we will now demonstrate $\hat{D}_s$ is effectively the matrix given in Eq.~\eqref{eq:LowSoft} that we set out to justify. In order to do this we have to evaluate the remaining terms:
\begin{equation}\begin{aligned}
\mathfrak{S} &\equiv - \sum_{\langle i j \rangle} \mathbf{t}_i \cdot \mathbf{t}_j U_{ij}\,, \\
\mathfrak{D}_W &\equiv \sum_{\langle i j \rangle} \mathbf{t}_{3i} \cdot \mathbf{t}_{3j} U_{ij}\,, \\
\mathfrak{D}_Z &\equiv - \sum_{\langle i j \rangle} \mathbf{t}_{Zi} \cdot \mathbf{t}_{Zj} U_{ij}\,.
\label{eq:ndlsfulldef2}
\end{aligned}\end{equation}
The form of each of these matrices can be evaluated by acting with them on the operators -- the unbroken operators in Eq.~\eqref{eq:UnbrokenOps} for $\mathfrak{S}$ and the broken operators in Eq.~\eqref{eq:BrokenOps} for $\mathfrak{D}_{W/Z}$ -- where the action of the gauge index operators is given by Eq.~\eqref{eq:gaugeindexopaction}. Doing this, we find:
\begin{equation}
\mathfrak{S} = \begin{bmatrix}
2-2i\pi & 1-i\pi \\
0 & i\pi - 1
\end{bmatrix}\,,
\label{eq:frakS}
\end{equation}
whilst
\begin{equation}\begin{aligned}
\mathfrak{D}_{W,1-8} &= {\rm diag} \left( 0, 0, 0, 0, -1, -1, -1, -1 \right)\,, \\
\mathfrak{D}_Z &= - c_W^4 \mathfrak{D}_W\,.
\label{eq:frakD}
\end{aligned}\end{equation}
Substituting these results into Eq.~\eqref{eq:ndlsfulldef}, we find:
\begin{equation}
\hat{D}_{s,1-8} = \begin{bmatrix}
s_W^2 \left[ 1 + G(\mu) \right] & s_W^2 \\
s_W c_W \left[ 1 + G(\mu) \right] & s_W c_W \\
s_W c_W \left[ 1 + G(\mu) \right] & s_W c_W \\
c_W^2 \left[ 1 + G(\mu) \right] & c_W^2 \\
s_W^2 \left[ 1 + H(\mu) \right] & s_W^2 I(\mu) \\
s_W c_W \left[ 1 + H(\mu) \right] & s_W c_W I(\mu) \\
s_W c_W \left[ 1 + H(\mu) \right] & s_W c_W I(\mu) \\
c_W^2 \left[ 1 + H(\mu) \right] & c_W^2 I(\mu)
\end{bmatrix}\,,
\label{eq:softmatrixfull}
\end{equation}
where we have defined:
\begin{equation}\begin{aligned}
G(\mu) \equiv& \frac{\alpha_W(\mu)}{2\pi} \ln \frac{m_W^2}{\mu^2} \left( 2-2i\pi \right)\,, \\
H(\mu) \equiv& \frac{\alpha_W(\mu)}{2\pi} \ln \frac{m_W^2}{\mu^2} \left( 1-2i\pi \right) \\
&+ c_W^4 \frac{\alpha_Z(\mu)}{2\pi} \ln \frac{m_Z^2}{\mu^2}\,, \\
I(\mu) \equiv& \frac{\alpha_W}{2\pi} \ln \frac{m_W^2}{\mu^2} (1-i\pi)\,.
\label{eq:softmatrixfullentries}
\end{aligned}\end{equation}
From the form of $\hat{D}_s$ given in Eq.~\eqref{eq:softmatrixfull}, we can again reduce this to a $2 \times 2$ matrix which maps onto $\Sigma_1$ and $\Sigma_1 - \Sigma_2$, exactly as we did for the tree-level low-scale matching. Doing this, the $2 \times 2$ matrix we obtain is exactly Eq.~\eqref{eq:LowSoft}, which we set out to justify.

\subsection*{Final State Graphs}

Finally we have the last contribution, which is the combination of final state collinear graphs as well as $C_3 + C_4$, as defined in Eq.~\eqref{eq:nrsoftcollinear}. As mentioned in the previous subsection, this calculation has already been performed in \cite{Chiu:2009ft}, and given that the form of $C_i$ is the same for our kinematics as it is for theirs, we take the result from their work. In that paper they calculated this diagonal contribution for all possible weak bosons. For our calculation we are only interested in a final state photon or $Z$, for which they give:
\begin{equation}\begin{aligned}
D_c^Z = &\frac{\alpha_2}{2\pi} \left[ F_W + f_S \left( \frac{m_Z^2}{m_W^2}, 1 \right) \right] \\
&+ \frac{1}{2} \delta \mathfrak{R}_Z + \tan \bar{\theta}_W \mathfrak{R}_{\gamma \to Z}\,, \\
D_c^{\gamma} = &\frac{\alpha_2}{2\pi} \left[ F_W + f_S \left( 0, 1 \right) \right] \\
&+ \frac{1}{2} \delta \mathfrak{R}_{\gamma} + \cot \bar{\theta}_W \mathfrak{R}_{Z \to \gamma}\,.
\label{eq:CFKMcollinear}
\end{aligned}\end{equation}
The various terms in these equations are outlined below. Nonetheless, once the full expressions are written out the analytic result for the terms in Eq.~\eqref{eq:LowColinearConsts2} can be extracted as the terms independent of $\ln \mu^2$. 

To begin with we have:
\begin{equation}\begin{aligned}
F_W \equiv& \ln \frac{m_W^2}{\mu^2} \ln \frac{s}{\mu^2} - \frac{1}{2} \ln^2 \frac{m_W^2}{\mu^2} \\
&- \ln \frac{m_W^2}{\mu^2} - \frac{5\pi^2}{12} + 1\,,
\end{aligned}\end{equation}
where note for our calculation $s = 4 m_{\chi}^2$. Next $f_S(w, z)$ is defined as:
\begin{equation}
f_S(w,z) \equiv \int_0^1 dx \frac{(2-x)}{x} \ln \frac{1-x+zx-wx(1-x)}{1-x}\,,
\end{equation}
such that an explicit calculation gives us
\begin{equation}\begin{aligned}
f_S \left( \frac{m_Z^2}{m_W^2}, 1 \right) &= 1.08355\,, \\
f_S \left( 0, 1 \right) &= \frac{\pi^2}{3} - 1\,.
\end{aligned}\end{equation}
Finally the $\mathfrak{R}$ contributions are defined by:\footnote{Note there is a typo in Eq.~B2 of \cite{Chiu:2009ft}, where $\mathfrak{R}_{\gamma \to Z}$ and $\mathfrak{R}_{Z \to \gamma}$ involved $\Pi^{\prime}$ rather than $\Pi$. The expressions stated here are the correct ones, and we thank Aneesh Manohar for confirming this and for  providing a numerical cross check on our results for these terms.}
\begin{equation}\begin{aligned}
\delta \mathfrak{R}_Z &\equiv \Pi^{\prime}_{ZZ}(m_Z^2)\,, \\
\delta \mathfrak{R}_\gamma &\equiv \Pi^{\prime}_{\gamma \gamma}(0)\,, \\
\mathfrak{R}_{\gamma \to Z} &\equiv \frac{1}{m_Z^2} \Pi_{Z\gamma} (m_Z^2)\,, \\
\mathfrak{R}_{Z \to \gamma} &\equiv -\frac{1}{m_Z^2} \Pi_{\gamma Z} (0)\,,
\end{aligned}\end{equation}
where $\Pi^{\prime} \equiv \partial \Pi(k^2)/\partial k^2$ and the various $\Pi$ functions are defined via the inverse of the transverse gauge boson propagator
\begin{equation}
-i \left( g_{\mu \nu} - \frac{k_{\mu} k_{\nu}}{k^2} \right) \begin{bmatrix} k^2 - m_Z^2 - \Pi_{ZZ}(k^2) & - \Pi_{Z\gamma}(k^2) \\ - \Pi_{\gamma Z}(k^2) & k^2 - \Pi_{\gamma \gamma}(k^2) \end{bmatrix}\,.
\end{equation}
The form of the $\Pi$ functions is not given explicitly in \cite{Chiu:2009ft}, but can be determined from the results of e.g. \cite{Denner:1991kt,Bardin:1999ak}. When doing so, there are two factors that must be accounted for. Firstly the $\Pi$ functions must be calculated in $\overline{\rm MS}$. This is because \cite{Chiu:2009ft} accounts for the residues explicitly in \eqref{eq:CFKMcollinear}. If we used the on-shell scheme for external particles, as we did for the high-scale matching, we would double count the contribution from the residues. Secondly we need to respect that the low-scale matching is performed above and below the electroweak scale, which means the $\Pi$ functions for the photon and $Z$ must be treated differently. Above the matching scale the $W$, $Z$, top and Higgs are dynamical degrees of freedom, but below it they are not. Light degrees of freedom like the photon, bottom quark or electron are dynamical above and below. This means for the $Z$ contributions, we need to include all degrees of freedom -- heavy and light -- in the loops, as the $Z$ itself does not propagate below the matching and the light fermions are offshell in these loops. For the photon contributions, however, only the heavy degrees of freedom should be included. Accounting for these factors, we arrive at the following:
\begin{eqnarray}
\delta \mathfrak{R}_Z &=& \frac{\alpha_2}{4\pi} \left[ \frac{5-10s_W^2+46s_W^4}{6c_W^2} \ln \frac{m_Z^2}{\mu_Z^2} \right. \nn
&&\hspace{0.67in}\left.+ 1.5077 - 9.92036 i \vphantom{\frac{m_Z^2}{\mu_Z^2}}\right]\,,\nn
\delta \mathfrak{R}_\gamma &=& \frac{\alpha_2}{4\pi} \left[ - \frac{11}{9} s_W^2 \ln \frac{m_Z^2}{\mu_Z^2} + 0.8257 \right]\,, \nn
\mathfrak{R}_{\gamma \to Z} &=& \frac{\alpha_2}{4\pi}\left[ - \frac{7s_W^2+34s_W^4}{6c_W^2 \tan \bar{\theta}_W} \ln \frac{m_Z^2}{\mu_Z^2} \right. \nn 
&&\hspace{0.47in}\left.+ 0.3678 - 2.2748 i \vphantom{\frac{m_Z^2}{\mu_Z^2}}\right]\,, \nn
\mathfrak{R}_{Z \to \gamma} &=& \frac{\alpha_2}{4\pi} \left[ 2 s_W c_W \ln \frac{m_Z^2}{\mu_Z^2} - 0.2099 \right]\,.
\end{eqnarray}
Analytic forms for the $\Pi$ functions are provided in App.~\ref{app:PiFunctions}, we do not provide the full expressions here as they are lengthy. In order to determine the numerical values above we have used the following:
\begin{equation}\begin{aligned}
m_Z &= 91.1876~{\rm GeV}\,, \\
m_W &= 80.385~{\rm GeV}\,, \\
m_t &= 173.21~{\rm GeV}\,, \\
m_H &= 125~{\rm GeV}\,, \\
m_b &= 4.18~{\rm GeV}\,, \\
m_c &= 1.275~{\rm GeV}\,, \\
m_{\tau} &= 1.77682~{\rm GeV}\,, \\
m_s &= m_d = m_u = m_{\mu} = m_e = 0~{\rm GeV}\,, \\
c_W &= m_W/m_Z\,.
\end{aligned}\end{equation}
This completes the list of ingredients for Eq.~\eqref{eq:CFKMcollinear}. Substituting them into that equation gives exactly the relevant terms in Eqs.~\eqref{eq:LowColinear}, \eqref{eq:LowColinearConsts1}, and \eqref{eq:LowColinearConsts2}, justifying the diagonal part of the low-scale matching.  Note that the results are insensitive to the precise values used for the $m_b$ and $m_c$ masses.

We have now justified each of the pieces making up the low-scale one-loop matching. All that remains is to cross check this result, which we turn to in the next appendix.
\newpage
\section{Consistency Check on the Low-Scale Matching}
\label{app:consistencylow}

In this appendix we provide a cross check on the low-scale one-loop matching calculation, much as we did for the high-scale result in App.~\ref{app:consistency}. Given that we already cross checked the high-scale result, we here make use of that to determine whether the $\ln \mu$ contributions at the low scale are correct. In order to do this, we take Eq.~\eqref{eq:Running} and turn off the running, which amounts to setting $\mu_{m_{\chi}} = \mu_Z \equiv \mu$. In detail we obtain:
\begin{equation}
\begin{bmatrix} C_{\pm}^X \vspace{0.1cm}\\ C^X_{0} \end{bmatrix} = e^{\hat{D}^X(\mu)} \begin{bmatrix} C_1(\mu) \\ C_2(\mu) \end{bmatrix}\,.
\label{eq:norunning}
\end{equation}
Now as we have the full one-loop result, the $\ln \mu$ dependence between these two terms must cancel at $\mathcal{O}(\alpha_2)$ for any $X$, which we will now demonstrate.

Before doing this in general, we first consider the simpler case where electroweak symmetry remains unbroken and we just have a $W^3 W^3$ final state. In this case, as in general, to capture all $\mu$ dependence at $\mathcal{O}(\alpha_2)$ we also need to account for the $\beta$-function. If SU(2)$_{\rm L}$ remains unbroken, however, this is just simply captured in:
\begin{equation}
\alpha_2(\mu) = \alpha_2(m_Z) + \alpha_2^2(m_Z)^2 \frac{b_0}{4\pi} \ln \frac{m_Z^2}{\mu^2}\,,
\label{eq:coup1loop}
\end{equation}
where $b_0 = (43-2 n_D)/6$, with $n_D$ the number of SM doublets. This follows directly from Eq.~\eqref{eq:coupling}. In the unbroken theory we can simply set $c_W = 1$ and $s_W = 0$, so if we do this and substitute our results from Eqs.~\eqref{eq:WilsonCoeff}, \eqref{eq:lowbreakdown}, \eqref{eq:LowSoft}, \eqref{eq:LowColinear}, \eqref{eq:LowColinearConsts1} into Eq.~\eqref{eq:norunning}, then we find:
\begin{equation}\begin{aligned}
C_{\pm}^{W^3} &= \frac{1}{m_{\chi}} \left( \frac{b_0}{4} + c_1^{W^3} - 1 \right) \ln \mu^2 + \mu {\rm -ind.}\,, \\
C_{0}^{W^3} &= \mu {\rm -ind.}\,, \\
\end{aligned}\end{equation}
Now we can calculate that $c_1^{W^3} = (2 n_D - 19)/24$, which taking $n_D=12$ exactly agrees with $c_1^Z$ in Eq.~\eqref{eq:LowColinearConsts1} when $c_W=1$ and $s_W=0$ as it must. Then recalling $b_0$ from above we see that both coefficients are then $\mu$ independent at this order, demonstrating the required consistency.

We now consider the same cross check in the full broken theory. The added complication here is that for our different final states, $\gamma \gamma$, $\gamma Z$, and $ZZ$, the coupling is actually $s_W^2 \alpha_2$, $s_W c_W \alpha_2$, and $c_W^2 \alpha_2$ respectively. As we work in $\overline{\rm MS}$, we need to account for the fact that $s_W$ and $c_W$ are functions $\mu$. Explicit calculation demonstrates that the running is only relevant for the consistency of $C_{\pm}^X$ -- the cancellation in $C_{0}^X$ is independent of the $\beta$-function at this order -- and in fact we find:
\begin{equation}\begin{aligned}
C_{\pm}^X &= \frac{1}{m_{\chi}} \left( \frac{b_0^{(X)}}{4} + \frac{1}{2} \sum_{i \in X} c_1^i - 1 \right) \ln \mu^2 + \mu {\rm -ind.}\,.
\label{eq:genlowcc}
\end{aligned}\end{equation}
To derive this we simply used Eq.~\eqref{eq:coup1loop}, with $b_0 \to b_0^{(X)}$, leaving us to derive the appropriate form of $b_0^{(X)}$ for $X = \gamma \gamma$, $\gamma Z$, $ZZ$. Firstly note that
\begin{equation}\begin{aligned}
s_W^2(\mu) &= \frac{\alpha_1(\mu)}{\alpha_1(\mu)+\alpha_2(\mu)}\,, \\
c_W^2(\mu) &= \frac{\alpha_2(\mu)}{\alpha_1(\mu)+\alpha_2(\mu)}\,,
\label{eq:swcwrun}
\end{aligned}\end{equation}
where $\alpha_1$ is the U(1)$_Y$ coupling. We can write a similar expression to Eq.~\eqref{eq:coup1loop} for $\alpha_1$, but this time we have $b_0^{(1)} = -41/6$. To avoid confusion we also now refer to the SU(2)$_{\rm L}$ $b_0$ as $b_0^{(2)}=19/6$.

Now for the case of two $Z$ bosons in the final state, the appropriate $\beta$-function is:
\begin{equation}
\beta_{ZZ} = \mu \frac{d}{d\mu} \left[ c_W^2 \alpha_2 \right]\,.
\end{equation}
Combining this with Eq.~\eqref{eq:swcwrun}, we conclude that:
\begin{equation}\begin{aligned}
b^{(ZZ)}_0 =& \left( s_W^2 + 1 \right) b_0^{(2)} - \frac{s_W^4}{c_W^2} b_0^{(1)} \\
=& \frac{19+22s_W^4}{6c_W^2}\,.
\end{aligned}\end{equation}
There is an additional factor of $c_W^2$ in this expression than if we were just calculating the $\beta$-function for $\alpha_Z$. The reason for this is that $b^{(ZZ)}_0$ is the appropriate replacement for $b_0$ in Eq.~\eqref{eq:coup1loop}, which represents the correction to $\alpha_2 = c_W^2 \alpha_Z$ not $\alpha_Z$. Substituting this into Eq.~\eqref{eq:genlowcc} along with the definition of $c_1^Z$ from Eq.~\eqref{eq:LowColinearConsts1} demonstrates consistency for the $ZZ$ case.

The case of two final state photons has to be treated differently, because of the fact our low-scale matching integrated out the electroweak degrees of freedom, which did not include the photon. This means we need to use a modified version of the SU(2)$_{\rm L}$ and U(1)$_Y$ couplings that only include the running due to the modes being removed. This amounts to accounting for the running from the Higgs, $W$ and $Z$ bosons, and the top quark, which we treat as an SU(2)$_{\rm L}$ singlet Dirac fermion to ensure it is entirely removed through the matching. Doing so, the SM $\beta$-functions now evaluate to $b_0^{(2) \prime} = 43/6$ and $b_0^{(1) \prime} = -35/18$. Repeating the same calculation as we used to determine $b^{(ZZ)}_0$, we find that:
\begin{equation}
b^{(\gamma \gamma)}_0 = \left(b_0^{(1) \prime} + b_0^{(2) \prime} \right) s_W^2 = \frac{47}{9} s_W^2\,.
\end{equation}
Again, substituting this into Eq.~\eqref{eq:genlowcc} shows that the two photon case is also consistent. The final case $\gamma Z$, but it is straightforward to see that in this case Eq.~\eqref{eq:genlowcc} breaks into two conditions that are satisfied if the $ZZ$ and $\gamma \gamma$ cases are, so this is not an independent cross check. 

As such, in the absence of running, all the $\mu$ dependence in our calculation vanishes at $\mathcal{O}(\alpha_2)$, as it must. But we emphasise that this is a non-trivial cross check, that involves all aspects of the calculation in the full broken theory.

\newpage
\section{Analytic Form of $\Pi$}
\label{app:PiFunctions}

Here we state the analytic expressions for the $\overline{\rm MS}$ electroweak $\Pi$ functions for photon and $Z$ boson, appropriate for the matching from SCET$_{\rm EW}$ to SCET$_{\gamma}$. These results can be determined using standard references, such as \cite{Denner:1991kt,Bardin:1999ak}. As the photon is a dynamical degree of freedom above and below the matching, we only need to consider loop diagrams involving electroweak modes that are integrated out through the matching. This simplifies the evaluation, and we have the following two functions:
\begin{equation}\begin{aligned}
\Pi^{\prime}_{\gamma \gamma}(0) = \frac{\alpha_2s_W^2}{4\pi} &\left\{ - \frac{16}{9} \ln \frac{\mu^2}{m_t^2} + 3 \ln \frac{\mu^2}{m_W^2} + \frac{2}{3} \right\}\,, \\
\Pi_{\gamma Z}(0) = \frac{\alpha_2s_W^2}{4\pi} &\left\{ \frac{2m_W^2}{s_W c_W} \ln \frac{\mu^2}{m_W^2} \right\}\,.
\end{aligned}\end{equation}
As the $Z$ itself is being integrated out, we need to include all relevant loops when calculating $\Pi_{Z \gamma}$ and $\Pi^{\prime}_{ZZ}$. In order to simplify the expressions, we firstly introduce the following expressions:
\begin{eqnarray}
\beta &\equiv &\sqrt{\frac{4m^2}{s}-1}\,,\;\;\; \xi \equiv \sqrt{1-\frac{4m^2}{s}}\,, \\
\lambda_{\pm} &\equiv &\frac{1}{2s} \left( s - m_2^2 + m_1^2 \pm \sqrt{(s-m_2^2+m_1^2)^2-4s (m_1^2-i\epsilon)} \right)\,. \nonumber
\end{eqnarray}
In terms of these we then define:
\begin{equation}\begin{aligned}
a(m_1,m_2) \equiv &1 + \frac{m_1^2}{m_2^2-m_1^2} \ln \frac{m_1^2}{m_2^2}\,, \\
b(s,m) \equiv &2 + i \beta \ln \left( \frac{\beta+i}{\beta-i} \right)\,, \\
b_2(s,m) \equiv &2 - \xi \ln \frac{1+\xi}{1-\xi} + i \pi \xi\,, \\
c(s,m) \equiv &- \frac{2m^2}{s^2\beta} \left( \frac{2\beta}{1+\beta^2} + i \ln \frac{\beta+i}{\beta-i} \right)\,, \\
c_2(s,m) \equiv &\frac{2m^2}{s^2 \xi} \left( \frac{2 \xi}{\xi^2-1} - \ln \frac{1+\xi}{1-\xi} \right)\,, \\
d(s,m_1,m_2) \equiv &2  + \lambda_+ \ln \left( \frac{\lambda_+-1}{\lambda_+} \right) - \ln \left( \lambda_+ -1 \right) \\
&+ \lambda_- \ln \left( \frac{\lambda_--1}{\lambda_-} \right) - \ln \left( \lambda_- -1 \right)\,, \\
e(s,m_1,m_2) \equiv &- \frac{1}{s} + \ln \left( \frac{\lambda_+-1}{\lambda_+} \right) \frac{\partial \lambda_+}{\partial s} \\
&+ \ln \left( \frac{\lambda_--1}{\lambda_-} \right) \frac{\partial \lambda_-}{\partial s}\,.
\end{aligned}\end{equation}

\begin{widetext}
We can now write out the full expressions:
\vspace{-2cm}
\begin{equation}\begin{aligned}
\Pi_{Z \gamma}(m_Z^2) = \frac{\alpha_2s_W^2}{4\pi} &\left\{ \frac{6-16s_W^2}{9c_W s_W} \left[ \frac{1}{3} m_Z^2 - m_Z^2 \ln \frac{\mu^2}{m_t^2} - (m_Z^2 + 2m_t^2) b(m_Z^2, m_t) \right] \right. \\
&+ \frac{3-4s_W^2}{9c_W s_W} \left[ \frac{1}{3} m_Z^2 - m_Z^2 \ln \frac{\mu^2}{m_b^2} - (m_Z^2 + 2m_b^2) b_2(m_Z^2, m_b)\right] \\
\vphantom{\Pi_{Z \gamma}(m_Z^2) = \frac{\alpha_2s_W^2}{4\pi}} \hspace{3cm} &+ \frac{6-16s_W^2}{9c_W s_W} \left[ \frac{1}{3} m_Z^2 - m_Z^2 \ln \frac{\mu^2}{m_c^2} - (m_Z^2 + 2m_c^2) b_2(m_Z^2, m_c) \right] \\
&+ \frac{1-4s_W^2}{3 c_W s_W} \left[ \frac{1}{3} m_Z^2 - m_Z^2 \ln \frac{\mu^2}{m_{\tau}^2} - (m_Z^2 + 2m_{\tau}^2) b_2(m_Z^2, m_{\tau}) \right] \\
&+ m_Z^2 \frac{16s_W^2-6}{3c_W s_W} \left[ \frac{5}{3} + i \pi + \ln \frac{\mu^2}{m_Z^2} \right] \\
&+ \frac{1}{3s_W c_W} \left\{ \left[ \left(9 c_W^2 + \frac{1}{2} \right) m_Z^2 + \left( 12 c_W^2 + 4 \right) m_W^2 \right] \left(\ln \frac{\mu^2}{m_W^2} + b(m_Z^2,m_W) \right) \right. \\
&\hspace{1.9cm}\left. \left.- (12 c_W^2 - 2) m_W^2 \ln \frac{\mu^2}{m_W^2} + \frac{1}{3} m_Z^2 \right\} \right\}\,,
\end{aligned}\end{equation}
\vspace{4cm}

\noindent and finally
\begin{equation}\begin{aligned}
\Pi^{\prime}_{ZZ}(m_Z^2) = \frac{\alpha_2s_W^2}{4\pi} &\left\{ 2 \left\{ \frac{9 - 24 s_W^2 + 32 s_W^4}{36 c_W^2 s_W^2} \left[ - 
\ln \frac{\mu^2}{m_t^2} - b(m_Z^2,m_t) \hspace{6.2cm} \right.\right.\right. \\
&\left.\left.- (m_Z^2+2m_t^2) c(m_Z^2,m_t) + \frac{1}{3} \right] + \frac{3}{4s_W^2c_W^2} m_t^2 c(m_Z^2,m_t) \right\} \\
&+ 2 \left\{ \frac{9 - 12 s_W^2 + 8 s_W^4}{36 c_W^2 s_W^2} \left[ - \ln \frac{\mu^2}{m_b^2} - b_2(m_Z^2,m_b)\right.\right. \\
&\left.\left.- (m_Z^2+2m_b^2) c_2(m_Z^2,m_b) + \frac{1}{3} \right] + \frac{3}{4s_W^2c_W^2} m_b^2 c_2(m_Z^2,m_b) \right\} \\
&+ 2 \left\{ \frac{9 - 24 s_W^2 + 32 s_W^4}{36 c_W^2 s_W^2} \left[ - \ln \frac{\mu^2}{m_c^2} - b_2(m_Z^2,m_c)\right.\right. \\
&\hspace{0.8cm}\left.\left.- (m_Z^2+2m_c^2) c_2(m_Z^2,m_c) + \frac{1}{3} \right] + \frac{3}{4s_W^2c_W^2} m_c^2 c_2(m_Z^2,m_c) \right\} \\
&+ \frac{2}{3} \left\{ \frac{1 - 4 s_W^2 + 8 s_W^4}{4 c_W^2 s_W^2} \left[ - \ln \frac{\mu^2}{m_{\tau}^2} - b_2(m_Z^2,m_{\tau})\right.\right. \\
&\hspace{0.8cm}\left.\left.- (m_Z^2+2m_{\tau}^2) c_2(m_Z^2,m_{\tau}) + \frac{1}{3} \right] + \frac{3}{4s_W^2c_W^2} m_{\tau}^2 c_2(m_Z^2,m_{\tau}) \right\} \\
&+  \frac{7-12s_W^2+16s_W^4}{3 s_W^2 c_W^2} \left[ - \frac{2}{3} - \ln \frac{\mu^2}{m_Z^2} - i \pi \right] \\
&+ \frac{1}{6s_W^2c_W^2} \left\{ \left( 18 c_W^4 + 2 c_W^2 - \frac{1}{2} \right) \left( \ln \frac{\mu^2}{m_W^2} + b(m_Z^2,m_W) \right) + \frac{1}{3} \left( 4 c_W^2-1 \right) \right. \\
&\hspace{1.9cm}\left.+ \left[ \left( 18 c_W^4 + 2 c_W^2 - \frac{1}{2} \right) m_Z^2 + \left( 24 c_W^4 + 16 c_W^2 - 10 \right) m_W^2 \right] c(m_Z^2,m_W) \right\}\\
&+ \frac{1}{12s_W^2c_W^2} \left\{ - \left( \ln \frac{\mu^2}{m_Z^2} + d(m_Z^2,m_Z,m_H) \right) + \left( 2 m_H^2 - 11 m_Z^2 \right) e(m_Z^2,m_Z,m_H) \right. \\
&\hspace{2.07cm} - \frac{(m_Z^2-m_H^2)^2}{m_Z^2} e(m_Z^2,m_Z,m_H) - \frac{2}{3} \\
&\hspace{2.02cm}\left. \left.+ \frac{(m_Z^2-m_H^2)^2}{m_Z^4} \left( \ln \frac{m_H^2}{m_Z^2} + d(m_Z^2,m_Z,m_H) - a(m_Z,m_H) \right) \right\} \right\}\,.
\end{aligned}\end{equation}
\end{widetext}

\bibliography{HeavyDMatOneLoop}

\end{document}